\documentclass[preprintnumbers,nofootinbib,noshowpacs,eqsecnum,prd]{revtex4}

\usepackage{graphicx}
\usepackage{amsmath,amsthm,amssymb,amsfonts}
\usepackage{mathrsfs,bbm}
\usepackage{color,hyperref}

\definecolor{lgray}{gray}{0.9} 		
\graphicspath{ {figures/} }

\makeatletter       

\renewcommand{\p@subsection}{}

\makeatother

\newtheorem{theorem}{Theorem}

\newcommand*{\eweakgroup}{\mbox{$SU(2)_L \times U(1)_Y$} }
\newcommand*{\emgroup}{\mbox{$U(1)_{em}$} }

\newcommand*{\unitmatrix}{\mathbbm{1}}

\newcommand*{\twomat}[1]{\underline{#1}}             
\newcommand*{\tvec}[1]{\boldsymbol{#1}}              
\newcommand*{\trans}{\mathrm{T}}                     
\newcommand*{\by}{\!\times\!}                        
\newcommand*{\im}[1]{\text{Im} {#1}}                        
\newcommand*{\re}[1]{\text{Re} {#1}}

\DeclareMathOperator{\trace}{tr}

\DeclareMathOperator{\diag}{diag}		

\begin{document}

\title{Stability and symmetry breaking in the general three-Higgs-doublet model}

\author{M. Maniatis}
    \email[E-mail: ]{MManiatis@ubiobio.cl}
\affiliation{Departamento de Ciencias B\'a{}sicas, 
Universidad del B\'i{}o B\'i{}o, Casilla 447, Chill\'a{}n, Chile.}
\author{O. Nachtmann}
    \email[E-mail: ]{O.Nachtmann@thphys.uni-heidelberg.de}
\affiliation{Institut f\"ur Theoretische Physik, Philosophenweg 16, 69120
Heidelberg, Germany}


\begin{abstract}
Stability, electroweak symmetry breaking, and the stationarity equations of
the general three-Higgs-doublet model (3HDM)
where all doublets carry the same hypercharge
 are discussed in detail. Employing
the bilinear formalism the study of the 3HDM potential turns out to be straightforward.
For the case that the potential leads to the physically relevant electroweak 
symmetry breaking we present explicit formulae for the masses of the 
physical Higgs bosons.

\end{abstract}


\maketitle


\section{Introduction}
\label{intro}

T.D.~Lee has shown decades ago that 
in the general
two-Higgs-doublet model~(THDM) CP violation is
possible in the Higgs sector~\cite{Lee:1973iz}.
Meanwhile a lot of effort has been spent to investigate the THDM; 
see for instance the review~\cite{Branco:2011iw} and references therein.
In particular, some progress could be made
employing the bilinear approach. The bilinears appear
naturally in the Higgs potential in any n-Higgs doublet model~(nHDM),
since only the gauge-invariant scalar products of the Higgs-boson
doublet fields may appear in the potential. The bilinear formalism was developed
in detail in~\cite{Nagel:2004sw,Maniatis:2006fs}
and independently in~\cite{Nishi:2006tg}.

Initiated by these works, many aspects of the THDM and the general
nHDM were considered within this formalism. 
For instance, CP-violation properties of the THDM were presented in~\cite{Nishi:2006tg, Maniatis:2007vn}. 
Different symmetries of the THDM and the general nHDM were  
considered in some detail employing bilinears; see for 
instance~\cite{Ma:2009ax, Grzadkowski:2010dj, Ferreira:2010yh, Nishi:2007nh, Ivanov:2010ww, Ivanov:2010wz,Keus:2013hya}.
Relations between vacua of different properties in multi-Higgs-doublet models
were derived in~\cite{Barroso:2006pa}.

In this work we will focus on the three-Higgs-doublet model~(3HDM).
Many of the properties of this model are direct 
generalizations of the THDM, but there appear also new aspects.
As we will see in detail, the space of Higgs-boson doublets does, in terms
of bilinears, not 
correspond to the forward light cone space, as in case
of the THDM~\cite{Maniatis:2006fs}, but to a certain subspace; see
\cite{Maniatis:2007vn,Nishi:2006tg,Ivanov:2012fp}.
Driven mainly by the quark- and neutrino mixing data, several 
3HDM's have been proposed; see for instance~\cite{Chen:2004rr,Lavoura:2007dw,Aranda:2012bv,Aranda:2013kq}

In an analogous way to the study of the THDM in~\cite{Maniatis:2006fs}
we will discuss in the following, in sections \ref{bilinears} to \ref{potentialEW},
stability, electroweak symmetry breaking, and the
stationarity  points of the potential for any 3HDM.
In section \ref{potentialEW} we discuss the potential after
symmetry breaking, section \ref{conclusion} presents our conclusion.
Throughout the study we will illustrate the general results by two simple illustrative 
3HDM examples.
In appendix \ref{appendixA} we give mathematical relations
concerning bilinears. In appendix \ref{appendixB} we discuss
an explicit non-trivial example of a 3HDM, based on
an $O(2) \times \mathbbm{Z}_2$ symmetry~\cite{Grimus:2008dr}.


\section{Bilinears}
\label{bilinears}

We consider the tree-level Higgs potential of models with
three Higgs-boson doublets
satisfying \eweakgroup electroweak
gauge symmetry. This is a generalization of 
the case of two Higgs-boson doublets which were
discussed in detail in~\cite{Maniatis:2006fs}.

We assume that all doublets 
carry hypercharge $y=+1/2$ and denote the
complex doublet fields by
\begin{equation}
\label{eq-doubldef}
\varphi_i(x) = 
\begin{pmatrix} 
\varphi^+_i(x) \\
 \varphi^0_i(x)
 \end{pmatrix}; \qquad i=1,2,3.
\end{equation}
In the most general \eweakgroup gauge invariant Higgs potential
the Higgs-boson doublets enter solely via products of the
following form:
\begin{equation}
\label{eq-potterms}
\varphi_i(x)^{\dagger}\varphi_j(x),
\qquad i,j \in \{1,2,3\}.
\end{equation}

It is convenient to discuss the properties of the Higgs potential such
as its stability and its stationary points in terms of
gauge invariant bilinears. 

First we introduce the $3 \times 2$~matrix of the Higgs-boson fields
in the following way,
\begin{equation} \label{2.3}
\phi = 
\begin{pmatrix} 
\varphi^+_1 & \varphi^0_1  \\
\varphi^+_2 & \varphi^0_2 \\
\varphi^+_3 & \varphi^0_3 
\end{pmatrix} = 
\begin{pmatrix} 
\varphi_1^\trans \\
\varphi_2^\trans \\
\varphi_3^\trans \\
\end{pmatrix} .
\end{equation}
We arrange all possible \eweakgroup invariant
scalar products into the hermitian $3\by 3$~matrix
\begin{gather}
\label{eq-kmat}
\twomat{K} =  \phi \phi^\dagger =
\begin{pmatrix}
  \varphi_1^{\dagger}\varphi_1 & \varphi_2^{\dagger}\varphi_1 & \varphi_3^{\dagger}\varphi_1 \\
  \varphi_1^{\dagger}\varphi_2 & \varphi_2^{\dagger}\varphi_2 & \varphi_3^{\dagger}\varphi_2 \\
  \varphi_1^{\dagger}\varphi_3 & \varphi_2^{\dagger}\varphi_3 & \varphi_3^{\dagger}\varphi_3   
\end{pmatrix}.
\end{gather}

A basis for the $3 \times 3$ matrices is given by
\begin{equation} \label{2.4a}
\lambda_\alpha, \quad \alpha = 0,1,\ldots,8
\end{equation}
where 
\begin{equation}\label{2.4b}
\lambda_0 = \sqrt{\frac{2}{3}} \unitmatrix_3
\end{equation}
and $\lambda_a$, $a=1,\ldots,8$, are the Gell-Mann matrices. Here and in
the following greek indices ($\alpha$, $\beta$, $\ldots$) run from 0 to 8 and 
latin indices ($a$, $b$, $\ldots$) from 1 to 8.
We have
\begin{equation} \label{2.4c}
\trace (\lambda_\alpha \lambda_\beta) = 2 \delta_{\alpha \beta},\qquad
\trace (\lambda_\alpha) = \sqrt{6}\; \delta_{\alpha 0}.
\end{equation}
The decomposition of~$\twomat{K}$~\eqref{eq-kmat} reads now
\begin{equation} \label{2.5}
\twomat{K} = \frac{1}{2} K_\alpha \lambda_\alpha
\end{equation}
where the real coefficients $K_\alpha$ are given by 
\begin{equation} \label{2.6}
K_\alpha = K_\alpha^* = \trace (\twomat{K} \lambda_\alpha).
\end{equation}
With the matrix $\twomat{K}$, as defined in terms of the doublets 
in~\eqref{eq-kmat}, as well as the decomposition~\eqref{2.5}, \eqref{2.6},
we immediately express the scalar products in terms of the bilinears,
\begin{alignat}{3} \label{phiK}
\nonumber
&\varphi_1^\dagger \varphi_1 = \frac{K_0}{\sqrt{6}} +\frac{K_3}{2} +\frac{K_8}{2\sqrt{3}},\qquad
&&\varphi_1^\dagger \varphi_2 = \frac{1}{2}\left( K_1 + i K_2 \right),\qquad
&&\varphi_1^\dagger \varphi_3 = \frac{1}{2}\left( K_4 + i K_5 \right),\\
&\varphi_2^\dagger \varphi_2 = \frac{K_0}{\sqrt{6}} -\frac{K_3}{2} +\frac{K_8}{2\sqrt{3}},\qquad
&&\varphi_2^\dagger \varphi_3 = \frac{1}{2}\left( K_6 + i K_7 \right),\qquad
&&\varphi_3^\dagger \varphi_3 = \frac{K_0}{\sqrt{6}} - \frac{K_8}{\sqrt{3}}.
\end{alignat}

In the following we shall frequently use also
\begin{equation} \label{2.10a}
\begin{split}
&K_+ = \sqrt{\frac{2}{3}} K_0 + \sqrt{\frac{1}{3}} K_8 = \varphi_1^\dagger \varphi_1 + \varphi_2^\dagger \varphi_2,\\
&K_- = \sqrt{\frac{1}{3}} K_0 - \sqrt{\frac{2}{3}} K_8 = \sqrt{2} \varphi_3^\dagger \varphi_3 .
\end{split}
\end{equation}
From \eqref{phiK} follows 
\begin{equation} \label{2.10b}
K_3 = \varphi_1^\dagger \varphi_1 - \varphi_2^\dagger \varphi_2,
\end{equation}
therefore, we have the inequalities
\begin{equation} \label{2.10c}
K_+ \ge |K_3| \ge 0,
\qquad
K_- \ge 0.
\end{equation}
Furthermore, we see from \eqref{2.10a} that $K_+ =0$ implies
$\varphi_1=\varphi_2=0$ which gives with \eqref{phiK}
\begin{equation} \label{2.10d}
K_1=K_2=\ldots = K_7 =0.
\end{equation}
Further discussion of the basis change from $\alpha=0,\ldots,8$ to
$+,1,\ldots,7,-$ is given in appendix \ref{appendixA}.

The matrix $\twomat{K}$~\eqref{eq-kmat} is positive semidefinite
which follows directly from its definition. This in turn gives
\begin{equation}
\begin{split}
\sqrt{\frac{3}{2}} K_0=\trace(\twomat{K})\ge 0,\qquad
\det(\twomat{K})\ge 0\;.
\end{split}
\end{equation}

The hermitian matrix $\twomat{K}$~\eqref{eq-kmat} is constructed from 
the Higgs field matrix, $\twomat{K}=\phi\phi^\dagger$. 
Therefore, the nine coefficients  
$K_\alpha$ of its decomposition~\eqref{2.5} are completely
fixed given the Higgs-boson fields.

Since the $3 \times 2$~matrix $\phi$ has trivially
rank smaller or equal 2,  this holds also for the matrix~$\twomat{K}$.
On the other hand, 
any hermitian $3 \times 3$ matrix with rank equal or smaller than~2
which clearly has then
vanishing determinant, $\det(\twomat{K})=0$,
determines the Higgs-boson fields
$\varphi_i$, $i=1,2,3$ uniquely, up to a gauge transformation.
This was shown in detail in~\cite{Maniatis:2006fs} in their theorem~5
for the general case of n-Higgs-boson doublets.
In appendix~\ref{appendixA} we show that the gauge orbits of the three
Higgs fields~\eqref{eq-doubldef} are characterised by the following
set in the 9-dimensional space of $(K_0, \ldots, K_8)$:
\begin{equation} \label{2.9}
\begin{split}
&K_0 \ge  0,\\
&(\trace(\twomat{K}))^2-\trace(\twomat{K}^2) = 
K_0^2 - \frac{1}{2} K_a K_a \ge 0,\\
&\det(\twomat{K}) = \frac{1}{12} G_{\alpha \beta \gamma} K_\alpha K_\beta K_\gamma = 0.
\end{split}
\end{equation}
Here $G_{\alpha \beta \gamma}$ are completely symmetric constants defined in~\eqref{A.25},
\eqref{A.26}.
That is, to every gauge orbit of the Higgs-boson fields
corresponds exactly one vector~$(K_\alpha)$ satisfying~\eqref{2.9} and vice versa.
The first two relations of~\eqref{2.9} are analogous to the {\em light cone} conditions
of the THDM; see (36) of~\cite{Maniatis:2006fs}. The determinant relation,
trilinear in the~$K_\alpha$, is specific for the 3HDM.
Further discussions of the matrices $\twomat{K}$ with rank 0, 1, 2
are presented in appendix \ref{appendixA}.

Based on the bilinears we shall
in the following discuss the potential, 
basis transformations, stability, minimization, and electroweak symmetry breaking  
of the general 3HDM.

\section{The 3HDM potential and basis transformations} \label{basis}

In terms of the bilinear coefficients, $K_0$, $K_a$, $a=1,...,8$ 
we can write the general
3HDM potential in the form
\begin{equation}
\label{eq-vdef}
V = 
\xi_0 K_0 + \xi_a K_a + \eta_{00} K_0^2 + 2 K_0 \eta_a K_a + K_a \eta_{ab} K_b,
\end{equation}
where the  54 parameters $\xi_0$, $\xi_a$, $\eta_{00}$, $\eta_a$
and \mbox{$\eta_{ab}=\eta_{ba}$} are real. The potential~\eqref{eq-vdef}
contains all possible linear and
quadratic terms of the bilinears -- corresponding to all gauge invariant
quadratic and quartic terms of the Higgs-boson doublets.
Terms higher than quadratic in the bilinears should not appear in the
potential with view
of renormalizability. Any constant term in the potential can be dropped
and therefore~\eqref{eq-vdef} is the most general 3HDM potential.
We also introduce the notation 
\begin{equation} \label{3.1b}
\tvec{K}=(K_1, \ldots, K_8)^\trans, \quad
\tvec{\xi}=(\xi_1, \ldots, \xi_8)^\trans, \quad
\tvec{\eta}=(\eta_1, \ldots, \eta_8)^\trans, \quad
E=(\eta_{ab}), \quad
(\tilde{E}_{\alpha \beta}) =
\begin{pmatrix} \eta_{00} & \eta_b \\ \eta_a & \eta_{ab} \end{pmatrix}.
\end{equation}
We can then write the potential~\eqref{eq-vdef} in the compact form
\begin{equation} \label{potKvec}
V = \xi_\alpha K_\alpha + K_\alpha \tilde{E}_{\alpha \beta} K_\beta .
\end{equation}

Let us now consider a change of basis of the Higgs-boson fields,
$\varphi_i(x) \rightarrow \varphi'_i(x)$, where
\begin{equation}
\label{eq-udef}
\begin{pmatrix} \varphi'_1(x)^\trans \\
                \varphi'_2(x)^\trans \\
                \varphi'_3(x)^\trans \\                
                \end{pmatrix}
= U  \begin{pmatrix} 
\varphi_1(x)^\trans \\
\varphi_2(x)^\trans \\
\varphi_3(x)^\trans
\end{pmatrix}, 
\end{equation}
with $U \in U(3)$ a $3 \times 3$ unitary transformation, 
that is, $U^\dagger U = \unitmatrix_3$. From~\eqref{eq-udef} we have
$\phi'(x)=U \phi(x)$, 
for the matrix~$\twomat{K}$~\eqref{eq-kmat} 
\begin{equation} \label{3.2a}
\twomat{K}'(x) = U \twomat{K}(x) U^\dagger ,
\end{equation}
and for the bilinears
\begin{equation} \label{3.3}
K_0'(x) =  K_0(x), \qquad K_a'(x) = R_{ab}(U) K_b(x).
\end{equation}
Here \mbox{$R_{ab}(U)$} is defined by
\begin{equation} \label{3.5}
U^\dagger \lambda_a U = R_{ab}(U)\,\lambda_b.
\end{equation}
The matrix~\mbox{$R(U)$} has the properties
\begin{equation}
\label{eq-rprodet}
R^\ast(U)=R(U),
\quad
R^\trans(U)\, R(U) = \unitmatrix_8,
\quad
\det R(U) = 1,
\end{equation}
that is, $R(U)\in SO(8)$.
But the $R(U)$ form only a subset of $SO(8)$.

For the bilinears a pure phase transformation, $U=\exp (i \alpha) \unitmatrix_3$,
plays no role. We shall, therefore, consider here only transformations~\eqref{eq-udef}
with $U \in SU(3)$. In the transformation of the bilinears~\eqref{3.3}
$R_{ab}(U)$ is then the $8 \times 8$ matrix corresponding to~$U$ in the
adjoint representation of~$SU(3)$.

The Higgs potential~\eqref{eq-vdef} remains unchanged under
the replacement~\eqref{3.3} if we perform an appropriate transformation
of the parameters
\begin{equation}
\label{eq-partrafo}
\begin{split}
\xi'_0 &= \xi_0, \qquad   \tvec{\xi}' = R(U)\,\tvec{\xi}, \\
\eta'_{00} &= \eta_{00}, \qquad  \tvec{\eta}' = R(U)\,\tvec{\eta}, \qquad
 E' = R(U)\,E\,R^\trans(U).
 \end{split}
\end{equation}
In the pure 3HDM potential, that is the model without fermions,
we can use~\eqref{eq-partrafo} to bring e.g. $\tvec{\xi}$ to a standard form.
Consider the hermitian matrix
\begin{equation}\label{3.7a}
\underline{\Lambda}_\xi = \xi_a \lambda_a .
\end{equation}
Applying a transformation $U \in SU(3)$ we get with \eqref{3.5}--\eqref{eq-partrafo}
\begin{equation} \label{3.7b}
U \underline{\Lambda}_\xi U^\dagger = R_{ba}(U) \xi_a \lambda_b =
\xi_b' \lambda_b \equiv \underline{\Lambda}_{\xi'}
\end{equation}
With a suitable transformation $U$ we can, therefore, diagonalise $\underline{\Lambda}_\xi$.
That is, we always can achieve the form
\begin{equation} \label{3.7c}
\underline{\Lambda}_{\xi'} = \xi_3' \lambda_3 + \xi_8' \lambda_8, \qquad
\tvec{\xi}' = \left( 0,0, \xi_3', 0, 0, 0, 0, \xi_8' \right)^\trans .
\end{equation}
The number of relevant parameters of the general 3HDM potential is, therefore,
\begin{equation} \label{3.7d}
54-6=48.
\end{equation}
Note that instead of $\tvec{\xi}$ we could have chosen $\tvec{\eta}$ in the above argument.
Note also the slick proof of \eqref{3.7c} and \eqref{3.7d} employing the bilinear formalism.

Let us remark on the basis transformations with respect to the 3-Higgs-doublet model.
In a realistic model we have to consider, besides the Higgs potential,
kinetic terms for the Higgs-boson doublet fields as well as
Yukawa terms which provide couplings of the Higgs-boson doublets to fermions.
Under a basis transformation, that is, a transformation of the 
Higgs-boson doublets of the form~\eqref{eq-udef}, or 
equivalently, in terms of the bilinears, a transformation of
the form~\eqref{3.3}, the kinetic terms of the Higgs doublets
will remain invariant. 
However, we emphasize that, in general, the Yukawa couplings are {\em not} invariant 
under such a change of basis. 

In order to illustrate the use of the bilinears 
we will consider two simple examples of
explicit 3HDM Higgs potentials,
\begin{align}
&\text{Example }I
&V_I = - \mu^2 \varphi_3^\dagger \varphi_3 + \lambda (\varphi_1^\dagger \varphi_1
+ \varphi_2^\dagger \varphi_2 + \varphi_3^\dagger \varphi_3)^2 .
\end{align}
Here $\mu^2>0$ is a parameter of dimension mass squared and $\lambda>0$ is dimensionless.
Employing~\eqref{phiK} we write this potential in terms 
of the bilinears as
\begin{equation} \label{Vexp}
V_I = -\frac{\mu^2}{\sqrt{6}} K_0 + \frac{\mu^2}{\sqrt{3}} K_8
+ \frac{3}{2} \lambda K_0^2 =
-\frac{\mu^2}{\sqrt{2}} K_- + \lambda (K_+ +\frac{1}{\sqrt{2}} K_-)^2.
\end{equation}
This corresponds to the general form~\eqref{eq-vdef} with parameters,
\begin{equation}
\label{exp-parameters}
\xi_0 = -\frac{\mu^2}{\sqrt{6}},  \qquad 
\tvec{\xi} = \mu^2 (0,0,0,0,0,0,0,\frac{1}{\sqrt{3}})^\trans, \qquad
\eta_{00} = \frac{3}{2} \lambda, \qquad \tvec{\eta} =0,\qquad E = 0.
\end{equation}
In the basis $+,1,\ldots,7,-$ (see \eqref{A33} to \eqref{A36}) this gives
for $\xi$ and $\tilde{E}$,
\begin{equation} \label{3.16a}
\begin{split}
&\xi_- = -\frac{\mu^2}{\sqrt{2}}, \quad \xi_+=\xi_1=\ldots=\xi_7=0, \\
&\tilde{E}_{++} = \lambda, \quad
\tilde{E}_{+-} =\tilde{E}_{-+} =  \frac{\lambda}{\sqrt{2}}, \quad
\tilde{E}_{--} = \frac{\lambda}{2}, \text{ and all other elements zero.}
\end{split}
\end{equation}
\begin{align} \label{3.17}  \nonumber
\text{Example }II \qquad &
V_{II} = 
m_1^2 \varphi_1^\dagger \varphi_1 +
m_2^2 \varphi_2^\dagger \varphi_2 
 - \mu^2  \varphi_3^\dagger \varphi_3
 + \lambda (\varphi_3^\dagger \varphi_3)^2 \\ \nonumber
 &  =
\frac{1}{2} m_1^2 \big(\sqrt{\frac{2}{3}} K_0 + \sqrt{\frac{1}{3}} K_8 + K_3 \big)
+
\frac{1}{2} m_2^2 \big(\sqrt{\frac{2}{3}} K_0 + \sqrt{\frac{1}{3}} K_8 - K_3 \big) \\ \nonumber
& -
\frac{1}{\sqrt{2}} \mu^2 \big(\sqrt{\frac{1}{3}} K_0 - \sqrt{\frac{2}{3}} K_8 \big)
+
\frac{1}{2} \lambda \big(\sqrt{\frac{1}{3}} K_0 - \sqrt{\frac{2}{3}} K_8 \big)^2\\
 &  =
\frac{1}{2} m_1^2 \big(K_+ + K_3 \big) 
+
\frac{1}{2} m_2^2 \big(K_+ - K_3 \big) 
-
\frac{1}{\sqrt{2}} \mu^2 K_- + \frac{1}{2} \lambda K_-^2, 
\end{align}
where we require
\begin{equation} \label{3.18}
m_1^2 >0, \quad m_2^2>0, \quad \mu^2>0, \quad \lambda>0.
\end{equation}
Here, in the basis $+,1,\ldots,7,-$ (see \eqref{A33} to \eqref{A36})
only the following elements of $\xi$ and $\tilde{E}$ are non zero
\begin{equation} \label{3.19}
\begin{split}
\xi_+ = \frac{1}{2} ( m_1^2 + m_2^2), \quad
\xi_3 = \frac{1}{2} ( m_1^2 - m_2^2), \quad
\xi_- = - \frac{1}{\sqrt{2}} \mu^2, \quad
\tilde{E}_{--} = \frac{1}{2} \lambda.
\end{split}
\end{equation}


\section{Stability of the 3HDM}
\label{stability}

Let us now analyse stability of the general 3HDM potential~(\ref{eq-vdef}),
given in terms of the bilinears $K_0$ and~$\tvec{K}$ on the domain
determined by~\eqref{2.9}.
This can be done in an analogous way to the THDM; see~\cite{Maniatis:2006fs}.  
The case $\sqrt{3/2} K_0=\varphi_1^\dagger \varphi_1 +
\varphi_2^\dagger \varphi_2 + \varphi_3^\dagger \varphi_3 = 0$
corresponds to vanishing Higgs-boson fields and $V=0$.
For \mbox{$K_0 > 0$} we define
\begin{equation}
\label{eq-ksde}
\tvec{k} = \frac{\tvec{K}}{K_0} = \left(\frac{K_a}{K_0}\right).
\end{equation}
Due to~\eqref{2.9} we have for~$\tvec{k}$ the domain~${\cal D}_{\tvec{k}}$:
\begin{equation} \label{4.1a}
\begin{split}
&2 - \tvec{k}^2 \ge 0,\\
&\det ( \sqrt{2/3} \unitmatrix_3 + k_a \lambda_a) = 0.
\end{split}
\end{equation}
The domain boundary, $\partial {\cal D}_{\tvec{k}}$, is characterised by 
\begin{equation} \label{4.1b}
2 - \tvec{k}^2 = 0.
\end{equation}

From~(\ref{eq-vdef}) and~(\ref{eq-ksde}) we obtain, for~\mbox{$K_0 > 0$},
$V=V_2+V_4$ with
\begin{align}
\label{eq-vk}
V_2 &= K_0\, J_2(\tvec{k}),&
J_2(\tvec{k}) &:= \xi_0 + \tvec{\xi}^\trans \tvec{k},\\
\label{eq-vk4}
V_4 &= K_0^2\, J_4(\tvec{k}),&
J_4(\tvec{k}) &:= \eta_{00} 
  + 2 \tvec{\eta}^\trans \tvec{k} + \tvec{k}^\trans E \tvec{k}
\end{align}
where we introduce the functions $J_2(\tvec{k})$ and $J_4(\tvec{k})$
on the domain~\eqref{4.1a}.

A stable potential means that it is bounded from below.
The stability is determined by the behaviour of~$V$ in the limit
\mbox{$K_0 \rightarrow \infty$}, that is, by the signs of
\mbox{$J_4(\tvec{k})$} and \mbox{$J_2(\tvec{k})$} in~\eqref{eq-vk}, \eqref{eq-vk4}.
For a model to be at least \emph{marginally} stable, the conditions
\begin{equation} \label{eq-margstab}
\begin{split}
  J_4(\tvec{k}) &> 0 \quad\text{or}\\
  J_4(\tvec{k}) &= 0 \quad\text{and}\quad  J_2(\tvec{k}) \ge 0
\end{split}
\end{equation}
for all~$\tvec{k} \in {\cal D}_{\tvec{k}}$, that is, all~$\tvec{k}$
satisfying~\eqref{4.1a}
are necessary and sufficient,
since this is equivalent to $V \ge 0$ for
$K_0 \rightarrow \infty$ in all possible allowed directions~$\tvec{k}$.
The more strict stability property $V \rightarrow \infty$ for
$K_0 \rightarrow \infty$ and any allowed~$\tvec{k}$ 
requires $V$ to be stable either in the strong or the weak sense.
For strong stability we require
\begin{equation} \label{4.5}
  J_4(\tvec{k}) > 0
\end{equation}
for all $\tvec{k} \in {\cal D}_{\tvec{k}}$; see~\eqref{4.1a}.
For stability in the weak sense we require for all 
$\tvec{k} \in {\cal D}_{\tvec{k}}$
\begin{equation} \label{4.6}
\begin{split}
  J_4(\tvec{k}) \ge & 0,\\
  J_2(\tvec{k}) > & 0 \text{ for all } \tvec{k} \text{ where } J_4(\tvec{k}) =0.
\end{split}
\end{equation}
To check that $J_4(\tvec{k})$ is positive (semi-)definite, it is sufficient to
consider its value for all stationary points 
on the domain~${\cal D}_{\tvec{k}}$.
This holds because
the global minimum of the continuous function \mbox{$J_4(\tvec{k})$}
is reached on the compact domain 
${\cal D}_{\tvec{k}}$, and the global minimum is among those
stationary points.

To obtain the stationary points of~$J_4(\tvec{k})$ in the interior of
the domain ${\cal D}_{\tvec{k}}$ we add to~$J_4(\tvec{k})$ the
second condition in~\eqref{4.1a} with a Lagrange multiplier~$u$.
The stationary points are then obtained from 
\begin{equation}  \label{4.7}
\begin{split}
& \nabla_{k_1,\ldots,k_8} \bigg[
J_4(\tvec{k})- u\cdot g(\tvec{k}) \bigg]  =0,\\
& g(\tvec{k}) = \det ( \sqrt{2/3} \unitmatrix_3 + k_a \lambda_a) = 0, \\
&2- \tvec{k}^2 > 0,
\end{split}
\end{equation}
provided the gradient matrix of the constraint equation has rank 1.
This can easily be checked. With $k_0=1$ we have from \eqref{A.24}
and \eqref{A.27}
\begin{equation} \label{4.9a}
\frac{1}{K_0} \twomat{K} = \frac{1}{2} k_\alpha \lambda_\alpha,
\end{equation}
\begin{equation} \label{4.9b}
g(\tvec{k}) = \det ( 2 \frac{\twomat{K}}{K_0}) =\frac{2}{3} G_{\alpha \beta \gamma} k_\alpha k_\beta k_\gamma,
\end{equation}
\begin{equation} \label{4.9c}
\frac{\partial g(\tvec{k})}{\partial k_a} =
2 G_{a \beta \gamma} k_\beta k_\gamma = 2 \frac{M_a}{K_0^2} .
\end{equation}
In our case the corresponding matrix $\twomat{M}= M_\alpha \lambda_\alpha/2$
has rank 1, see \eqref{A.20}, therefore, $M_0$ and at least one
element $M_a$ with $a \in \{1,\ldots,8\}$ have to be non zero.
That is, $\big(\partial g(\tvec{k})/\partial k_a \big)$ has rank 1 as required.

For the stationary points on the boundary~$\partial {\cal D}_{\tvec{k}}$
we have two constraints, see \eqref{4.1a}, \eqref{4.1b},
\begin{equation} \label{4.9d}
\begin{split}
g_1(\tvec{k}) = \det ( \sqrt{2/3} \unitmatrix_3 + k_a \lambda_a) = 0, \qquad
g_2(\tvec{k}) = 2 - k_a k_a =0.
\end{split}
\end{equation}
Here the gradient matrix reads
\begin{equation} \label{4.9e}
\begin{pmatrix}
\frac{\partial g_1(\tvec{k})}{\partial k_a}\\ 
\frac{\partial g_2(\tvec{k})}{\partial k_a}
\end{pmatrix}
=
\begin{pmatrix}
2 \frac{M_a}{K_0^2} \\
- 2 k_a
\end{pmatrix}.
\end{equation}
For $\twomat{K}$ as in \eqref{4.9a} but now of rank 1
we find from \eqref{A.20}, $\twomat{M}=0$ and, therefore, $M_a=0$,
$a=1, \ldots, 8$. Thus, the gradient matrix has here only rank 1
and not the required rank 2 for the application of the 
Lagrange multiplier method in an analogous way to \eqref{4.7}.
We turn, therefore, to the parametrization of the rank 1 matrices
$\twomat{K}$ of \eqref{A10a} to \eqref{A10c}:
\begin{equation} \label{4.9f}
\frac{\twomat{K}}{K_0} = \sqrt{\frac{3}{2}} \tvec{w} \tvec{w}^\dagger
\end{equation}
where 
\begin{equation} \label{4.9g}
\tvec{w}^\dagger \tvec{w} = 1.
\end{equation}
This gives, according to \eqref{2.6} and \eqref{eq-ksde}, \eqref{eq-vk4},
\begin{equation} \label{4.9h}
\begin{split}
&k_a \equiv k_a(\tvec{w}^\dagger, \tvec{w}) = \trace \big( \frac{1}{K_0} \twomat{K} \lambda_a \big)
=  \sqrt{\frac{3}{2}} \tvec{w}^\dagger \lambda_a \tvec{w},\\
& J_4(\tvec{k}) \equiv J_4(\tvec{w}^\dagger, \tvec{w}) = 
\eta_{00} + 2 \eta_a \sqrt{\frac{3}{2}}  \tvec{w}^\dagger \lambda_a \tvec{w} +
\frac{3}{2} \big(\tvec{w}^\dagger \lambda_a \tvec{w} \big) E_{ab} \big(\tvec{w}^\dagger \lambda_b \tvec{w} \big).
\end{split}
\end{equation}
Now we can determine the stationary points of $J_4(\tvec{w}^\dagger, \tvec{w})$
subject to the constraint \eqref{4.9g}. With $u$ a Lagrange
multiplier we get
\begin{equation} \label{4.9i}
\begin{split}
&\nabla_{\tvec{w}^\dagger} \big[ J_4(\tvec{w}^\dagger, \tvec{w})  - u (\tvec{w}^\dagger \tvec{w} - 1) \big] =0,\\
& \tvec{w}^\dagger \tvec{w} - 1 =0 .
\end{split}
\end{equation}
Here the gradient matrix of the constraint is of rank 1 as required and we get explicitly
\begin{equation} \label{4.10}
\begin{split}
& \big[ \sqrt{6} \eta_a \lambda_a + 
	3 E_{ab} \big(\tvec{w}^\dagger \lambda_b \tvec{w} \big) \lambda_a -u \big] \tvec{w} = 0,\\
& \tvec{w}^\dagger \tvec{w} - 1 =0 .
\end{split}
\end{equation}
All stationary points obtained from~\eqref{4.7} 
and \eqref{4.10} have to fulfill the condition
$J_4(\tvec{k})>0$ for stability in the strong sense.
If for all stationary points we have $J_4(\tvec{k})\ge 0$, then 
for every solution $\tvec{k}$ with 
$J_4(\tvec{k})=0$ we have to have 
$J_2(\tvec{k})>0$ for stability in the weak sense,
or at least
$J_2(\tvec{k})=0$ for {\em marginal} stability.
If none of these conditions is fulfilled, that 
is, if we find at least one stationary direction~$\tvec{k}$ 
with $J_4(\tvec{k})<0$ or $J_4(\tvec{k})=0$ but 
$J_2(\tvec{k})<0$, the
potential is unstable.

In our explicit example $I$, with $V_I$ from \eqref{Vexp},
the functions $J_2(\tvec{k})$ and $J_4(\tvec{k})$
read
\begin{equation}
J_2(\tvec{k}) = \left(-\frac{1}{\sqrt{6}} + \frac{k_8}{\sqrt{3}} \right) \mu^2,
\qquad
J_4(\tvec{k}) = \frac{3}{2} \lambda.
\end{equation}
Obviously, $J_4(\tvec{k})$ is always positive for $\lambda>0$ in any direction $\tvec{k}$,
therefore, the potential is stable in the strong sense. That is, stability is here guarantied by
the quartic terms of the potential alone.

For example $II$ from \eqref{3.17}, \eqref{3.18}, we get
\begin{equation} \label{4.13}
V_2 = \frac{1}{2} m_1^2 (K_+ + K_3) + \frac{1}{2} m_2^2 (K_+ - K_3) - \frac{1}{\sqrt{2}} \mu^2 K_-,
\qquad
V_4 = \frac{1}{2} \lambda K_-^2.
\end{equation}
We have $V_4>0$ for $K_- > 0$ but $V_4=0$ for $K_-=0$.
Thus, we have to investigate $V_2$ for $K_-=0$:
\begin{equation} \label{4.14}
V_2 \bigg|_{K_-=0} = \frac{1}{2} m_1^2 (K_+ + K_3) + \frac{1}{2} m_2^2 (K_+ - K_3) .
\end{equation}
Due to \eqref{2.10c} and \eqref{2.10d} we have
\begin{equation} \label{4.15}
V_2 \bigg|_{K_-=0} \ge 0
\end{equation}
where $V_2 \bigg|_{K_-=0} = 0$ only holds if 
\begin{equation} \label{4.16}
K_+ +K_3 =0, \qquad K_+ -K_3 =0,
\end{equation}
that is, for $K_+=0$. But this implies $\twomat{K}=0$.
Thus, the potential $V_{II}$ from \eqref{3.17}, \eqref{3.18} is stable in the weak sense.


\section{Electroweak symmetry breaking of the 3HDM}

Suppose now that the 3HDM potential is stable, that is, bounded from below.
Then the global minimum will be among the stationary points of~$V$.
In the following 
the different types of minima with respect to electroweak symmetry
breaking are discussed and the corresponding stationarity
equations are presented.

As we have discussed in section~\ref{bilinears}, 
the space of the
Higgs-boson doublets is determined, up to electroweak gauge transformations,
by the space of the hermitian $3 \times 3$ matrices $\twomat{K}$ with 
rank smaller or equal~2. Since the rank of the matrix  $\twomat{K}$
is equal to the rank of the Higgs-boson field matrix~$\phi$~\eqref{2.3}
we can distinguish the different types of minima with respect to
electroweak symmetry breaking as follows.
At the global minimum, that is, the vacuum configuration, we
write the $3 \times 2$~matrix of the Higgs-boson fields as
\begin{equation}
\label{5.1}
\langle \phi  \rangle= 
\begin{pmatrix} 
v^+_1 & v^0_1  \\
v^+_2 & v^0_2  \\
v^+_3 & v^0_3  \\
\end{pmatrix}.
\end{equation}
In the case this matrix has rank 2, we cannot, by a \eweakgroup 
transformation, achieve a form with all 
charged components~$v^+_i$, $i=1,2,3$ vanishing.
This means that the full \eweakgroup is broken. 
In case we have at the minimum a matrix 
$\langle \phi \rangle$ with rank one, we can, by a
\eweakgroup transformation, achieve a form with all charged 
components~$v_i^+$ vanishing. The unbroken~$U(1)$ gauge group can then be
identified with the electromagnetic gauge group. Therefore,
a minimum with rank one corresponds to the electroweak-symmetry
breaking \eweakgroup~$\rightarrow$~\emgroup\!\!.
Eventually, a vanishing matrix at the minimum, $\langle \phi  \rangle =0$,
corresponds to an unbroken electroweak symmetry. 
Of course, only a minimum with a partially broken electroweak symmetry
is physically acceptable.

We study now the matrix~$\twomat{K}_v$ corresponding 
to $\langle \phi  \rangle$~\eqref{5.1}
\begin{equation} \label{5.2}
\twomat{K}_v = \langle \phi \rangle \langle \phi \rangle^\dagger = 
\frac{1}{2} K_{v \alpha} \lambda_\alpha .
\end{equation}
For an acceptable vacuum~$\langle \phi  \rangle$, $\twomat{K}_v$
must have rank~1. From~\eqref{A.14} we see that $\twomat{K}_v$
has rank~1 and is positive semidefinite if and only if
\begin{equation} \label{5.3}
\begin{split}
&\trace \twomat{K}_v = \sqrt{\frac{3}{2}} K_{v 0} > 0,\\
& 2 K_{v 0}^2 - K_{v a} K_{v a} = 0,\\
&\det (\twomat{K}_v ) = 0.
\end{split}
\end{equation}
By a suitable~$U(3)$ transformation~\eqref{eq-udef} we can bring
the vacuum value~$\langle \phi \rangle$ of rank~1 to the form
\begin{equation} \label{5.4}
\langle \phi \rangle =
\begin{pmatrix}
0 & 0 \\
0 & 0 \\
0 & v_0/\sqrt{2}
\end{pmatrix}, \qquad v_0 >0.
\end{equation}
In a realistic model $v_0$ must be the usual Higgs-boson
vacuum expectation value,
\begin{equation} \label{5.5}
v_0 \approx 246~\text{GeV}.
\end{equation}
With~\eqref{5.4} we get in this basis a particularly simple form for
$\twomat{K}_v$ respectively $K_{v \alpha}$:
\begin{equation} \label{5.6}
\begin{split}
&\twomat{K}_v = 
\frac{1}{2}
\begin{pmatrix}
0 & 0 & 0 \\
0 & 0 & 0 \\
0 & 0 & v_0^2
\end{pmatrix} = \frac{1}{2} K_{v \alpha} \lambda_\alpha,\\
&\left( K_{v \alpha} \right) = 
\frac{v_0^2}{\sqrt{6}}
\begin{pmatrix}
1, & 0, & \ldots, & 0, &-\sqrt{2}
\end{pmatrix}^\trans\\
& K_{v+} =0, \quad K_{v-}= \frac{1}{\sqrt{2}} v_0^2.
\end{split}
\end{equation}

Another possible choice for the vacuum expectation value, obtainable by 
a suitable transformation \eqref{eq-udef} from \eqref{5.4} is
\begin{equation} \label{5.6a}
\langle \phi \rangle =
\begin{pmatrix}
0 & v_0/\sqrt{2} \\
0 & 0 \\
0 & 0
\end{pmatrix}, \qquad v_0 >0.
\end{equation}
Here we get
\begin{equation} \label{5.6b}
\begin{split}
\twomat{K}_v = &
\frac{1}{2}
\begin{pmatrix}
v_0^2 & 0 & 0 \\
0 & 0 & 0 \\
0 & 0 & 0
\end{pmatrix},\\
\left( K_{v \alpha} \right) = & v_0^2
\begin{pmatrix}
\frac{1}{\sqrt{6}}, & 0, & 0, & \frac{1}{2}, & 0, & 0, & 0, & 0, & \frac{1}{2\sqrt{3}}
\end{pmatrix}^\trans.
\end{split}
\end{equation}

In the cases where~$\langle \phi \rangle$ of \eqref{5.1} has rank~2
or rank~0 also the matrix~$\twomat{K}_v$, \eqref{5.2}, has rank~2 or zero,
respectively. The corresponding conditions for~$\twomat{K}_v$ are given explicitly
in~\eqref{A.13} and \eqref{A.15}, respectively. We can, therefore, summarise  our findings
for the vacuum values to  a given potential~$V$ as follows.

Let~$\langle \phi \rangle$ be the vacuum expectation value of the Higgs-boson field
matrix to a given, stable, potential~$V$ and $\twomat{K}_v = \langle \phi \rangle 
\langle \phi \rangle^\dagger = K_{v \alpha} \lambda_\alpha / 2$.
The gauge symmetry \eweakgroup is fully broken by the vacuum if and only if
\begin{equation} \label{5.7}
K_{v 0}>0, \qquad  2 K_{v 0}^2 - K_{v a} K_{v a} >0.
\end{equation}
We have the breaking \eweakgroup $\to$ \emgroup if and only if
\begin{equation} \label{5.8}
K_{v 0}>0, \qquad  2 K_{v 0}^2 - K_{v a} K_{v a} =0.
\end{equation}
We have no breaking of \eweakgroup if and only if
\begin{equation} \label{5.9}
K_{v \alpha} =0.
\end{equation}
Of course, we always have
\begin{equation} \label{5.10}
\det \twomat{K}_v = \frac{1}{12} G_{\alpha \beta \gamma} K_{v \alpha} K_{v \beta} K_{v \gamma} = 0
\end{equation}
with $G_{\alpha \beta \gamma}$ defined in~\eqref{A.25}.


\section{Stationary points}
\label{stationarity}

Following the study of stability and electroweak symmetry breaking
in the last two sections we shall now present the 
stationarity equations.
We suppose again that the potential is stable. Then the global minimum 
is among the stationary points of~$V$.

We classify the stationary points by the rank of the stationarity matrix~$\twomat{K}$.
In the following we use the conditions for~$\twomat{K}$ having 
rank~0, 1, 2, or 3 as given in appendix~\ref{appendixA}; see \eqref{A.12} -- \eqref{A.15}.

The matrix $\twomat{K}=0$, respectively $K_\alpha=0$, $\alpha=0,\ldots,8$,
always corresponds to a stationary point of~$V$ with value~$V(K_\alpha)=0$.

All stationarity matrices $\twomat{K}= K_\alpha \lambda_\alpha /2$ of rank~2 are
obtained from the following system of equations where~$u$ is a 
Lagrange multiplier:
\begin{equation} \label{5.12}
\begin{split}
& \nabla_{K_0,\ldots,K_8} \bigg[ V(K_0,\ldots,K_8) - 
u\; 
\det(\twomat{K})
\bigg] =0, \\
&2 K_0^2 - K_a K_a >0,\\
&\det(\twomat{K})=0,\\
& K_0 >0.
\end{split}
\end{equation}
Explicitly we get here, using \eqref{potKvec},
\begin{equation} \label{6.3}
\begin{split}
&\xi_\alpha + 2 \tilde{E}_{\alpha \beta} K_\beta 
- \frac{u}{4} G_{\alpha \beta \gamma} K_\beta K_\gamma =0,\\
&(3 \delta_{\alpha 0} \delta_{\beta 0} - \delta_{\alpha \beta} ) K_\alpha K_\beta > 0,\\
&G_{\alpha \beta \gamma} K_\alpha K_\beta K_\gamma =0,\\
&K_0 >0.
\end{split}
\end{equation}
The gradient matrix of the constraint is given by (see \eqref{A.24} and \eqref{A.27})
\begin{equation} \label{6.2a}
\big(\nabla_{K_\alpha} \det(\twomat{K})\big) = \big(\nabla_{K_\alpha} \big( \frac{1}{12} G_{\alpha' \beta \gamma} K_{\alpha'} K_\beta K_\gamma \big) \big)
=
\big( \frac{1}{4} G_{\alpha \beta \gamma} K_\beta K_\gamma \big) = \big( \frac{1}{4} M_\alpha \big)
\end{equation}
and has rank 1 as required. This holds since for $\twomat{K}$ of rank 2,
$\twomat{M}$ has rank 1, as we see from \eqref{A.20}, implying,
for instance, $M_0 = \trace (\twomat{M} \lambda_0) > 0$.

For the stationarity matrices $\twomat{K} = K_\alpha \lambda_\alpha /2$
of rank~1 we cannot use the Lagrange multiplier method in an analogous way to \eqref{5.12}. The two constraints
for rank 1, $g_1=2 K_0^2 - K_a K_a =0$ and $g_2=\det(\twomat{K}) =0$ yield a
gradient matrix 
of rank 1 for $\twomat{K}$ of rank 1. However, the Langrange multiplier method
requires that this matrix has rank 2 in this case.
The stationarity equations of the potential $V(\twomat{K})$ for $\twomat{K}$ of rank 1 follow
from a parametrization of the matrix $\twomat{K}$ of rank 1, as given in \eqref{A10c}:
\begin{equation} \label{Kw1}
\twomat{K}(K_0, \tvec{w}^\dagger, \tvec{w} ) = K_0 \sqrt{\frac{3}{2}} \tvec{w} \tvec{w}^\dagger
\end{equation}
where 
\begin{equation} \label{6.4}
K_0 > 0, \qquad \tvec{w}^\dagger \tvec{w} - 1 =0 .
\end{equation}
This gives
\begin{equation} \label{6.5}
K_\alpha(K_0, \tvec{w}^\dagger, \tvec{w} ) = 
\trace\big( \twomat{K}(K_0, \tvec{w}^\dagger, \tvec{w} )  \lambda_\alpha \big) =
K_0 \sqrt{\frac{3}{2}} \tvec{w}^\dagger \lambda_\alpha \tvec{w}.
\end{equation}

Taking into account the constraint \eqref{6.4} with a Lagrange multiplier $u$
we have to determine the stationary points of
\begin{equation}
V( K_\alpha(K_0, \tvec{w}^\dagger, \tvec{w} )) - u ( \tvec{w}^\dagger  \tvec{w} -1 ) 
\end{equation}
under variation of $K_0$, and $\tvec{w}^\dagger$, $\tvec{w}$. The gradient matrix
of the constraint has rank 1 as required and we get with \eqref{potKvec} the following
system of equations
\begin{equation} \label{6.6}
\begin{split}
&\bigg[ K_0 \sqrt{\frac{3}{2}} \xi_\alpha \lambda_\alpha +
3 K_0^2 \tilde{E}_{\alpha \beta}(\tvec{w}^\dagger \lambda_\beta \tvec{w}) \lambda_\alpha - u \bigg] \tvec{w}=0,\\
&\sqrt{\frac{3}{2}} \xi_\alpha (\tvec{w}^\dagger \lambda_\alpha \tvec{w}) +
3 K_0(\tvec{w}^\dagger \lambda_\alpha \tvec{w}) \tilde{E}_{\alpha \beta} (\tvec{w}^\dagger \lambda_\beta \tvec{w}) =0,\\
& \tvec{w}^\dagger \tvec{w} -1 = 0,\\
& K_0 >0.
\end{split}
\end{equation}

The stationarity matrix~$\twomat{K}=K_\alpha \lambda_\alpha /2$ with the
lowest value of $V(K_0,\ldots,K_8)$ gives the global minimum
$\twomat{K}_v$ of the potential.
Note that in general there may 
be degenerate global minima with the same potential value.
Systems of equations of the
kind~\eqref{6.3}, \eqref{6.6} can be solved via
the Groebner-basis approach or homotopy continuation;
see for instance~\cite{Maniatis:2006jd, Maniatis:2012ex}.


\section{The potential after symmetry breaking}
\label{potentialEW}

In this section we discuss the potential after symmetry breaking
and the procedure to calculate the physical Higgs-boson
masses and self couplings in the 3HDM.
We will assume that the potential is stable and leads to the desired electroweak
symmetry breaking \eweakgroup $\to$ \emgroup. In particular,
the global minimum is then a solution of the set of 
equations~\eqref{6.6}.
In this case we can, in the unitary gauge, by an electroweak gauge transformation
and a $U(3)$ rotation \eqref{eq-udef}
always achieve the form \eqref{5.4} for the vacuum expectation value 
of the Higgs-field matrix.
For the original Higgs fields expressed in terms of the physical fields we get then
\begin{equation} \label{scunitary}
\varphi_{1/2}(x) = 
  \begin{pmatrix} H_{1/2}^+(x)\\ \frac{1}{\sqrt{2}} \left( H_{1/2}^0(x)+ i A_{1/2}^0(x) \right)  \end{pmatrix},
\qquad
\varphi_3(x) = \frac{1}{\sqrt{2}}
  \begin{pmatrix} 0\\ v_0 + h_0(x) \end{pmatrix},
\end{equation}
with $v_0$ real and positive, neutral fields  $H_1^0(x)$, $A_1^0(x)$, $H_2^0(x)$, $A_2^0(x)$, $h_0(x)$,
as well as the complex charged fields $H_1^+(x)$ and $H_2^+(x)$.
The negatively charged
Higgs-boson fields are defined by $H_{1/2}^-(x) = \left(H_{1/2}^+(x)\right)^\dagger$.
Thus, we have in the 3HDM the following physical fields
\begin{equation} \label{fields-ph}
\begin{split}
&\text{five neutral fields:} \quad H_1^0(x), A_1^0(x), H_2^0(x), A_2^0(x),  h_0(x)\\
&\text{two charged fields:} \quad H_1^+(x), H_2^+(x).
\end{split}
\end{equation}
In general, however, the physical fields of definite mass are linear combinations of
the fields in~\eqref{fields-ph}.
Obviously, the 3 original complex doublets of any 3HDM, corresponding to 12 real degrees of freedom,
yield 5 real fields and 2 complex fields, with the 3 remaining degrees of freedom
absorbed via the mechanism of electroweak symmetry breaking. 

Throughout this section we shall work in a basis where $\twomat{K}_v$ has the
form \eqref{5.6}. Representing $\twomat{K}$ as in \eqref{Kw1} we get
\begin{equation} \label{7.3}
\twomat{K}_v = \frac{v_0^2}{2} \tvec{e}_3 \tvec{e}_3^\dagger, \qquad \tvec{w} = \tvec{e}_3,
\end{equation}
where $\tvec{e}_1$, $\tvec{e}_2$, $\tvec{e}_3$ are the three-dimensional Cartesian
unit vectors. That is, $K_0 = v_0^2/\sqrt{6}$ and $\tvec{w}= \tvec{e}_3$ must be
solutions of \eqref{6.6}. We get
\begin{equation} \label{7.3a}
\tvec{e}_3^\dagger \lambda_\alpha \tvec{e}_3 = \sqrt{\frac{2}{3}}
( \delta_{\alpha 0} - \sqrt{2} \delta_{\alpha 8})
\end{equation}
and define for $\alpha= 0, \ldots 8$
\begin{equation} \label{7.4}
\begin{split}
\zeta_\alpha & = \xi_\alpha + 2 \tilde{E}_{\alpha \beta} K_{v \beta} 
= \xi_\alpha + 2 \tilde{E}_{\alpha -} K_{v -},\\
\zeta_+ & = \xi_+ + 2 \tilde{E}_{+-} K_{v -},\\
\zeta_- & = \xi_- + 2 \tilde{E}_{--} K_{v -}
\end{split}
\end{equation}
where the $\pm$ components are defined in \eqref{A33} ff.
Inserting all this in \eqref{6.6} with $\tvec{w}=\tvec{e}_3$ we get from
the first equation there 
\begin{equation} \label{7.5a}
\big[\frac{v_0^2}{2} \zeta_\alpha \lambda_\alpha - u \big] \tvec{e}_3 = 0\\
\end{equation}
that is,
\begin{equation} \label{7.5b}
\frac{v_0^2}{2}\big(\tvec{e}_1( \zeta_4-i \zeta_5) + \tvec{e}_2(\zeta_6-i\zeta_7) \big)
+ \tvec{e}_3 \big( \frac{v_0^2}{\sqrt{2}} \zeta_- - u \big) = 0.
\end{equation}
The $\zeta_\alpha$ are all real, therefore, we have from \eqref{7.5b}
\begin{equation} \label{7.5c}
\zeta_4= \zeta_5 = \zeta_6=\zeta_7 =0, \qquad \frac{v_0^2}{\sqrt{2}} \zeta_- = u .
\end{equation}
The second equation of \eqref{6.6} gives
\begin{equation}
\zeta_-=0
\end{equation}
and thus, from \eqref{7.5c}, $u=0$.

To summarise, in the basis where $\langle \phi \rangle$ and $\twomat{K}_v$ have the
forms \eqref{5.4} and \eqref{5.6}, respectively, we have
\begin{equation} \label{7.7}
\zeta_\alpha = 0 \quad \text{for } \alpha = 4,5,6,7,- .
\end{equation}

The next task is to expand $\phi$, $\twomat{K}$, and $V$ in terms 
of the physical fields \eqref{fields-ph}. For $\phi$ we write
\begin{equation} \label{7.8}
\begin{split}
&\phi(x) = \langle \phi \rangle + \phi^{(1)}(x),\\
&\phi^{(1)}(x)= \langle \phi \rangle \frac{h_0(x)}{v_0} + \phi'^{(1)}(x),\\
&\phi'^{(1)}(x) =
\begin{pmatrix}
H_1^+(x) & \frac{1}{\sqrt{2}} ( H_1^0(x) + i A_1^0(x))\\
H_2^+(x) & \frac{1}{\sqrt{2}} ( H_2^0(x) + i A_2^0(x))\\
0 & 0
\end{pmatrix}.
\end{split}
\end{equation}
From this we get for $\twomat{K}(x)$ and $K_\alpha(x)$ the following
with $\twomat{K}_v$ and $K_{v \alpha}$ given in \eqref{5.6}
\begin{equation} \label{7.9}
\twomat{K}(x) = \twomat{K}_v + \twomat{K}^{(1)}(x) + \twomat{K}^{(2)}(x),
\end{equation}
\begin{equation} \label{7.10}
\twomat{K}^{(1)}(x) = \frac{2 h_0(x)}{v_0}\twomat{K}_v + \twomat{K}'^{(1)}(x),
\end{equation}
\begin{equation} \label{7.11}
\twomat{K}'^{(1)}(x) = \frac{v_0}{2} \bigg(
H_1^0(x) \lambda_4 - A_1^0(x) \lambda_5 + H_2^0(x) \lambda_6 - A_2^0(x) \lambda_7 \bigg),
\end{equation}
\begin{equation} \label{7.12}
\twomat{K}^{(2)}(x) = \phi^{(1)}(x)\phi^{(1)\dagger}(x) .
\end{equation}
\begin{equation} \label{7.13}
\begin{split}
&\big(K_\alpha^{(1)}(x)\big) = v_0 \bigg(
\sqrt{\frac{2}{3}} h_0(x), 0, 0, 0, H_1^0(x), -A_1^0(x),  H_2^0(x), -A_2^0(x), -\frac{2}{\sqrt{3}}h_0(x) \bigg)^\trans,
\\
&K_+^{(1)}(x) =0, \qquad K_-^{(1)}(x) = \sqrt{2} v_0 h_0(x),
\end{split}
\end{equation}
\begin{equation} \label{7.14}
\begin{split}
&K_0^{(2)}(x) = \sqrt{\frac{2}{3}} 
\bigg[
H_1^-(x) H_1^+(x) + H_2^-(x) H_2^+(x) 
+ \frac{1}{2}
\big(
\left(H_1^0(x)\right)^2 + \left(A_1^0(x)\right)^2 +
\left(H_2^0(x)\right)^2 + \left(A_2^0(x)\right)^2 +
\left(h_0(x)\right)^2 
\big)
\bigg],\\
&K_1^{(2)}(x) =  H_1^+(x) H_2^-(x) + H_1^-(x)H_2^+(x) + H_1^0(x) H_2^0(x) + A_1^0(x) A_2^0(x),\\
&K_2^{(2)}(x) =  i ( H_1^+(x)H_2^-(x) - H_1^-(x)H_2^+(x)) + H_1^0(x)A_2^0(x)- A_1^0(x)H_2^0(x),\\
&K_3^{(2)}(x) =  H_1^-(x)H_1^+(x) - H_2^-(x)H_2^+(x) + \frac{1}{2}
\big(
\left(H_1^0(x)\right)^2 + \left(A_1^0(x)\right)^2 -
\left(H_2^0(x)\right)^2 - \left(A_2^0(x)\right)^2
\big),\\
&K_4^{(2)}(x) =   H_1^0(x) h_0(x),\\
&K_5^{(2)}(x) =  - A_1^0(x) h_0(x),\\
&K_6^{(2)}(x) =   H_2^0(x) h_0(x),\\
&K_7^{(2)}(x) =  - A_2^0(x) h_0(x),\\
&K_8^{(2)}(x) =  \frac{1}{\sqrt{3}} 
\bigg[
H_1^-(x)H_1^+(x) + H_2^-(x)H_2^+(x) + 
+ \frac{1}{2}
\big(
\left(H_1^0(x)\right)^2 + \left(A_1^0(x)\right)^2 +
\left(H_2^0(x)\right)^2 + \left(A_2^0(x)\right)^2
\big) -
\left(h_0(x)\right)^2 
\bigg],\\
&K_+^{(2)}(x) =  
H_1^-(x)H_1^+(x) + H_2^-(x)H_2^+(x) 
+ \frac{1}{2}
\big(
\left(H_1^0(x)\right)^2 + \left(A_1^0(x)\right)^2 +
\left(H_2^0(x)\right)^2 + \left(A_2^0(x)\right)^2 
\big),\\
&K_-^{(2)}(x) =  \frac{1}{\sqrt{2}} \left(h_0(x)\right)^2.
\end{split}
\end{equation}
For the potential we have the following expansion in the order of 
the physical fields
\begin{equation} \label{7.15}
V = V^{(0)} + V^{(1)} + V^{(2)} + V^{(3)} + V^{(4)}.
\end{equation}
\begin{equation} \label{7.16}
\begin{split}
V^{(0)} = & K_{v \alpha} \xi_\alpha + K_{v \alpha} \tilde{E}_{\alpha \beta} K_{v \beta},\\
V^{(1)} = & K_{\alpha}^{(1)}(x) \xi_\alpha + 2 K_{\alpha}^{(1)}(x) \tilde{E}_{\alpha \beta} K_{v \beta},\\
V^{(2)} = & K_{\alpha}^{(2)}(x) \xi_\alpha + 2 K_{\alpha}^{(2)}(x) \tilde{E}_{\alpha \beta} K_{v \beta}
+ K_{\alpha}^{(1)}(x) \tilde{E}_{\alpha \beta} K_{\beta}^{(1)}(x),\\
V^{(3)} = & 2 K_{\alpha}^{(2)}(x) \tilde{E}_{\alpha \beta} K_{\beta}^{(1)}(x),\\
V^{(4)} = & K_{\alpha}^{(2)}(x) \tilde{E}_{\alpha \beta} K_{\beta}^{(2)}(x).
\end{split}
\end{equation}
We shall now discuss $V^{(0)}$, $V^{(1)}$,  and $V^{(2)}$,
where it is convenient to use the basis $(+,1,\ldots,7,-)$; see \eqref{A33} ff.
For $V^{(0)}$ we find with \eqref{5.6}, \eqref{7.4}, and \eqref{7.7},
\begin{equation} \label{7.17}
V^{(0)} =  K_{v -} \big( \xi_- + \tilde{E}_{--} K_{v -} \big)
= \frac{1}{2} K_{v-} \big( \xi_- + \zeta_-\big) 
= \frac{1}{2} K_{v-} \xi_- =
\frac{v_0^2}{2 \sqrt{6}} \big( \xi_0 - \sqrt{2} \xi_8 \big).
\end{equation}
For $V^{(1)}$ we get from  \eqref{5.6}, \eqref{7.4}, \eqref{7.7}, and \eqref{7.13},
\begin{equation} \label{7.18}
V^{(1)} = K_+^{(1)}(x) \zeta_+ + \sum_{\alpha=1}^7 K_\alpha^{(1)}(x) \zeta_\alpha + K_-^{(1)}(x) \zeta_- = 0.
\end{equation}
This must be so, since we are expanding around the global minimum.
From $V^{(2)}$ we get the mass matrices squared for the charged and neutral physical fields:
\begin{equation} \label{7.19}
V^{(2)} = \begin{pmatrix} H_1^-(x),& H_2^-(x) \end{pmatrix}   {\mathscr M}_{\text ch}^2
\begin{pmatrix} H_1^+(x)\\ H_2^+(x) \end{pmatrix}
+
\begin{pmatrix} H_1^0(x),& A_1^0(x),& H_2^0(x),& A_2^0(x),& h_0(x) \end{pmatrix}
\frac{1}{2} {\mathscr M}_{\text n}^2 
\begin{pmatrix} H_1^0(x)\\ A_1^0(x)\\ H_2^0(x)\\ A_2^0(x)\\ h_0(x) \end{pmatrix}.
\end{equation}
With \eqref{5.6}, \eqref{7.7}, \eqref{7.13}, and \eqref{7.14} we get
\begin{equation} \label{7.20}
{\mathscr M}_{\text ch}^2 = 
\begin{pmatrix}
\zeta_+ + \zeta_3 & \zeta_1 - i \zeta_2\\
\zeta_1 + i \zeta_2 & \zeta_+ - \zeta_3 
\end{pmatrix},
\end{equation}
\begin{equation} \label{7.21}
{\mathscr M}_{\text n}^2 =
\begin{pmatrix}
\zeta_+ + \zeta_3 + 2 v_0^2 \tilde{E}_{44} 
&
-2 v_0^2 \tilde{E}_{45} 
&
\zeta_1 + 2 v_0^2 \tilde{E}_{46} 
&
\zeta_2 -2 v_0^2 \tilde{E}_{47}
&
2 \sqrt{2} v_0^2 \tilde{E}_{4-}
\\
-2 v_0^2 \tilde{E}_{54}
&
\zeta_+ + \zeta_3+2 v_0^2 \tilde{E}_{55}
& 
-\zeta_2 -2 v_0^2 \tilde{E}_{56}
&
\zeta_1 + 2 v_0^2 \tilde{E}_{57}
&
-2 \sqrt{2}v_0^2 \tilde{E}_{5-}
\\
\zeta_1 + 2 v_0^2 \tilde{E}_{64}
&
-\zeta_2 - 2 v_0^2 \tilde{E}_{65}
& 
\zeta_+-\zeta_3 +2 v_0^2 \tilde{E}_{66}
&
-2 v_0^2 \tilde{E}_{67}
&
2 \sqrt{2}v_0^2 \tilde{E}_{6-}
\\
\zeta_2 - 2 v_0^2 \tilde{E}_{74}
&
\zeta_1 + 2 v_0^2 \tilde{E}_{75}
& 
-2 v_0^2 \tilde{E}_{76}
&
\zeta_+ - \zeta_3   +2 v_0^2 \tilde{E}_{77}
&
-2 \sqrt{2}v_0^2 \tilde{E}_{7-}
\\
2 \sqrt{2} v_0^2 \tilde{E}_{-4}
&
-2 \sqrt{2}v_0^2 \tilde{E}_{-5}
& 
2 \sqrt{2} v_0^2 \tilde{E}_{-6}
&
-2 \sqrt{2} v_0^2 \tilde{E}_{-7}
&
4v_0^2 \tilde{E}_{--}
\end{pmatrix}.
\end{equation}
Note that from \eqref{5.6}, \eqref{7.4}, and \eqref{7.7}
we get
\begin{equation} \label{7.24a}
\sqrt{2} v_0^2 \tilde{E}_{\alpha -} = -\xi_\alpha \text{ for } \alpha =4,5,6,7,-.
\end{equation}

Since we are expanding around the global minimum we must have that
$V^{(0)}$ is below or at most equal to $V(K_\alpha=0)=0$. From \eqref{7.17}
this implies
\begin{equation} \label{7.22}
V^{(0)} =
\frac{1}{2\sqrt{2}} v_0^2 \xi_- =
\frac{1}{2\sqrt{6}} v_0^2 (\xi_0 - \sqrt{2} \xi_8) \le 0.
\end{equation}
Furthermore, the mass squared matrices 
${\mathscr M}_{\text ch}^2$ and
${\mathscr M}_{\text n}^2$ must be positive semidefinite. This
implies, for instance, from \eqref{7.20} that we must have
\begin{equation}
\zeta_+ \ge \sqrt{\zeta_1^2 + \zeta_2^2 + \zeta_3^2}.
\end{equation}
To obtain the physical Higgs bosons of definite mass, the 
matrices \eqref{7.20} and \eqref{7.21} have to be diagonalised.
We note that the field $h_0(x)$ is a mass eigenstate if
\begin{equation}
\tilde{E}_{\alpha -} = 0 \qquad \text{for } \alpha = 4,5,6.7.
\end{equation}
In this case $h_0(x)$ is what is called {\em aligned} 
with the vacuum expectation value (see for instance \cite{Pich:2009sp})
and its mass squared is given by
\begin{equation} \label{7.27a}
m_{h_0}^2 = 4 v_0^2 \tilde{E}_{--}= -2 \sqrt{2}\xi_-= - \frac{8}{v_0^2} V^{(0)}.
\end{equation}

In our example $I$, the 3HDM Higgs potential~\eqref{Vexp},
we find stationary points for vanishing fields, corresponding
to an unbroken EW symmetry, from the set~\eqref{5.12}
we get no solution with $K_0>0$,  and 
from the set~\eqref{6.6} we get one solution with
\begin{equation} \label{7.24}
\tvec{w} = \tvec{e}_3, \qquad K_0 = \frac{1}{\sqrt{6}} \frac{\mu^2}{\lambda} .
\end{equation}
The corresponding
potential value is $V^{(0)}= -1/4 \cdot(\mu^2)^2/\lambda$ and is  the
deepest stationary point and therefore the global minimum. 
From~\eqref{5.6} we see that the global minimum
corresponds to a vacuum expectation value $v_0= \sqrt{\mu^2/\lambda}$.
For the mass matrices we get from \eqref{7.20} and \eqref{7.21}
\begin{equation} \label{7.25}
{\mathscr M}_{\text ch}^2=  \diag ( \lambda v_0^2, \lambda v_0^2), \quad
{\mathscr M}_{\text n}^2 = \diag ( \lambda v_0^2, \lambda v_0^2, \lambda v_0^2, \lambda v_0^2, 2 \lambda v_0^2).
\end{equation}

Turning to example $II$, $V_{II}$ of \eqref{3.17}, we have
as stationary points the trivial one, $K_\alpha=0$ with $V_{II}(0)=0$,
no stationary point from \eqref{5.12} and one point from \eqref{6.6}.
The latter is obtained again for
\begin{equation} \label{7.26}
\tvec{w} = \tvec{e}_3, \qquad K_0 = \frac{1}{\sqrt{6}} \frac{\mu^2}{\lambda} .
\end{equation}
Here we get from \eqref{5.6}, \eqref{7.20}, and \eqref{7.21}
\begin{equation} \label{7.27}
\begin{split}
&v_0^2 = \frac{\mu^2}{\lambda}, \qquad
\begin{pmatrix} K_{v +},& K_{v1},& \ldots &, K_{v7},&K_{v-} \end{pmatrix}
=
\begin{pmatrix} 0,& \ldots,& 0,& \frac{1}{\sqrt{2}} v_0^2 \end{pmatrix},\\
&{\mathscr M}_{\text ch}^2=  \diag ( m_1^2, m_2^2), \qquad
{\mathscr M}_{\text n}^2 = \diag ( m_1^2, m_1^2, m_2^2, m_2^2, 2 \lambda v_0^2).
\end{split}
\end{equation}
This simple example shows that the squared masses of the charged 
physical Higgs bosons need not be degenerate in a  3HDM having 
the correct electroweak symmetry breaking.
In appendix \ref{appendixB} a nontrivial example of a 3HDM is
discussed.


\section{Conclusion}
\label{conclusion}

The three-Higgs-doublet model has been studied as
a generalization of the THDM. 
Stability, electroweak symmetry breaking, and the types of
stationary points of the potential have been investigated.
Explicit sets of equations have been presented which allow to determine
the stability of any 3HDM and, in case of a stable potential,
to find the global minimum or the degenerate global minima
in case the potential has such.
For the  case that the 3HDM has the physically relevant electroweak
symmetry breaking $ \eweakgroup \to \emgroup$ we have given
explicit expressions for the mass squared matrices of
the charged and neutral physical Higgs bosons.
The use of bilinears
turns out to be very helpful: in particular,
irrelevant gauge degrees of freedom are avoided and the degree of the
polynomial equations which are to be solved is reduced in this formalism.
In general, the sets of equations which determine stability
and the stationary points are rather involved. However,
approaches like the Groebner-basis approach or homotopy
continuation may be applied to solve these systems of equations
in an efficient way.
This has been demonstrated for a 3HDM based on a 
$O(2) \times \mathbbm{Z}_2$ symmetry.

\begin{acknowledgments}
The work of M.M. was supported, in part, by Fondecyt (Chile)
Grant No. 1140568.
\end{acknowledgments}

\appendix

\section{Properties of the matrix $\twomat{K}$}
\label{appendixA}

Here we want to discuss the properties of the matrix~$\twomat{K}$~\eqref{eq-kmat} 
with respect to its rank.

First we note that the $3\times 3$ matrix $\twomat{K}$ is hermitian and 
positive semidefinite. Hence, we can, by a unitary
transformation, diagonalise this matrix,
 \begin{equation} \label{A.1}
 U \twomat{K} U^\dagger =
 \begin{pmatrix}
 \kappa_1 & 0 & 0 \\
 0 & \kappa_2 & 0 \\
 0 & 0 & \kappa_3
 \end{pmatrix},
 \end{equation}
with all $\kappa_i\ge 0$.
In particular, we have,
 \begin{equation} \label{A.2}
 \begin{split}
 &\trace(\twomat{K}) = \kappa_1+\kappa_2+\kappa_3,\\
  & (\trace(\twomat{K}))^2-\trace(\twomat{K}^2) =
 2 \kappa_1 \kappa_2 + 2 \kappa_2 \kappa_3 + 2 \kappa_1 \kappa_3,\\
&\det( \twomat{K}) = \kappa_1 \kappa_2 \kappa_3.
\end{split}
\end{equation}
Employing the properties of the Gell-Mann matrices~\eqref{2.4c}
we can write the second trace condition in the form 
 \begin{equation} \label{A.3}
 (\trace(\twomat{K}))^2-\trace(\twomat{K}^2) =
 K_0^2 - \frac{1}{2} K_a K_a.
 \end{equation}

Suppose now that the matrix~$\twomat{K}$ has rank~3, then,
we have to have for all three $\kappa_i$
\begin{equation} \label{A.4}
 \kappa_i>0.
 \end{equation}
 It follows immediately from \eqref{A.2}
 \begin{equation} \label{A.5}
 \trace(\twomat{K})>0, \quad
  (\trace(\twomat{K}))^2-\trace(\twomat{K}^2)>0, \quad \det(\twomat{K}) >0.
 \end{equation}
 If, for the reverse, we have for a hermitian matrix $\twomat{K}$ the conditions~\eqref{A.5} fulfilled,
 then, using \eqref{A.2} we find that we must have all $\kappa_i>0$.
 That is, $\twomat{K}$ has
 rank~3 and is positive definite.

Suppose the matrix~$\twomat{K}$ has rank~2, then,
without loss of generality, we can assume
\begin{equation} \label{A.6}
 \kappa_1>0, \quad \kappa_2>0, \quad \kappa_3=0.
\end{equation}
 It follows immediately from \eqref{A.2} that
 \begin{equation} \label{A.7}
 \trace(\twomat{K})>0, \quad
  (\trace(\twomat{K}))^2-\trace(\twomat{K}^2)>0, \quad \det(\twomat{K}) =0.
 \end{equation}
 If, for the reverse, we have for a hermitian matrix $\twomat{K}$ the 
 conditions~\eqref{A.7} fulfilled,
 then, from the last equation in~\eqref{A.2} at least one $\kappa_i=0$. Without loss of generality we can 
 suppose $\kappa_3=0$. We have then
 \begin{equation} \label{A.8}
 \begin{split}
 &\trace(\twomat{K}) = \kappa_1 + \kappa_2  >0,\\
 &  (\trace(\twomat{K}))^2-\trace(\twomat{K}^2) =  2  \kappa_1 \kappa_2 >0
 \end{split}
 \end{equation}
 which implies $\kappa_1>0$ and $\kappa_2>0$. That is, $\twomat{K}$ has
 rank~2 and is positive semidefinite.

Another way to characterise the positive semidefinite matrices of rank 2
is as follows. We set
\begin{equation} \label{A8a}
\kappa_1 = \sqrt{\frac{3}{2}} K_0 \sin^2 (\chi), \qquad
\kappa_2 = \sqrt{\frac{3}{2}} K_0 \cos^2 (\chi), \qquad
\kappa_3 = 0, \qquad 
K_0 >0, \quad 0 < \chi \le \frac{\pi}{4}.
\end{equation}
Let $\tvec{w}_1$ and $\tvec{w}_2$ be orthonormal eigenvectors of $\twomat{K}$
to $\kappa_1$ and $\kappa_2$, respectively, then we have
\begin{equation} \label{A8b}
\twomat{K} = K_0 \sqrt{\frac{3}{2}} \bigg(
\sin^2 (\chi) \tvec{w}_1 \tvec{w}_1^\dagger 
+
\cos^2 (\chi) \tvec{w}_2 \tvec{w}_2^\dagger 
\bigg),
\end{equation}
where
\begin{equation} \label{A8c}
\tvec{w}_i^\dagger \tvec{w}_j = \delta_{ij} .
\end{equation}
For $0 < \chi < \pi/4$ the $\tvec{w}_i$ are fixed up to phases,
for $\chi = \pi/4$ we may make arbitrary $U(2)$ rotations of $\tvec{w}_1$
and $\tvec{w}_2$. Clearly, every positive semidefinite matrix
$\twomat{K}$ of the form \eqref{A8b} has rank 2 and every positive
semidefinte matrix $\twomat{K}$ of rank 2 can be written in the form \eqref{A8b}.

Now, let us suppose the matrix~$\twomat{K}$ has rank~1, then,
without loss of generality, we can assume
\begin{equation} \label{A.9}
 \kappa_1>0, \quad \kappa_2=0, \quad \kappa_3=0.
\end{equation}
 It follows immediately from \eqref{A.2}
 \begin{equation} \label{A.10}
  \trace(\twomat{K})>0, \quad
  (\trace(\twomat{K}))^2-\trace(\twomat{K}^2)=0, \qquad \det(\twomat{K}) =0.
 \end{equation}
On the other hand, having the conditions \eqref{A.10} for a hermitian matrix $\twomat{K}$
fulfilled, employing~\eqref{A.2},
the determinant condition requires that at least one $\kappa_i$ vanishes,
for instance $\kappa_3=0$ without loss of generality.
Then the second condition requires that another eigenvalue has
to vanish, for instance $\kappa_2=0$. Eventually, the
first condition then dictates that the remaining $\kappa_1>0$. 
Hence, $\twomat{K}$ has rank~1 and is positive semidefinite.

Let now $\tvec{w}$ be eigenvector of $\twomat{K}$ to the eigenvalue 
$\kappa_1$ with 
\begin{equation} \label{A10a}
\tvec{w}^\dagger \tvec{w} = 1.
\end{equation}
We set
\begin{equation} \label{A10b}
\kappa_1 = K_0 \sqrt{\frac{3}{2}}, \qquad K_0>0.
\end{equation}
Then, any positive semidefinite matrix $\twomat{K}$ of rank 1
can be represented as 
\begin{equation} \label{A10c}
\twomat{K} = K_0 \sqrt{\frac{3}{2}} \tvec{w} \tvec{w}^\dagger .
\end{equation}
Conversely, any matrix of the form \eqref{A10c} with $K_0>0$ and 
$\tvec{w}^\dagger \tvec{w} = 1$ is a positive semidefinite matrix of rank 1.
Clearly, $\tvec{w}$ is fixed up to a phase transformation.

Finally, suppose the matrix~$\twomat{K}$ has rank~0,
then, clearly, all $\kappa_i$ have to vanish, 
corresponding to 
\begin{equation} \label{A.11}
 \trace(\twomat{K})   = 0, \quad
   (\trace(\twomat{K}))^2-\trace(\twomat{K}^2) =0, \quad
 \det(\twomat{K}) =0.
 \end{equation}
Vice versa, starting with the conditions \eqref{A.11} for a hermitian matrix $\twomat{K}$,
the determinant condition requires that one eigenvalue, for instance $\kappa_3=0$ has to vanish, 
the second condition in turn requires that another, say $\kappa_2=0$, 
and the first trace condition that also the third $\kappa_1=0$.
This means $\twomat{K}=0$.
 Therefore, we have shown the following theorem.
 \begin{theorem}: Let $\twomat{K}=K_\alpha \lambda_\alpha /2$ be a hermitian matrix.
 $\twomat{K}$ has rank~3 and is positive definite if and only if
 \begin{equation} \label{A.12}
 \begin{split}
 &\trace(\twomat{K}) = \sqrt{\frac{3}{2}} K_0  > 0,\\
 & 2 K_0^2 - K_a K_a >0,\\
 &\det(\twomat{K}) >0.
 \end{split}
 \end{equation}
 
 $\twomat{K}$ has rank~2 and is positive semidefinite if and only if
 \begin{equation} \label{A.13}
 \begin{split}
 &\trace(\twomat{K}) = \sqrt{\frac{3}{2}} K_0  > 0,\\
 & 2 K_0^2 - K_a K_a >0,\\
 &\det(\twomat{K}) =0.
 \end{split}
 \end{equation}
 
 $\twomat{K}$ has rank~1 and is positive semidefinite if and only if
 \begin{equation} \label{A.14}
 \begin{split}
 &\trace(\twomat{K}) = \sqrt{\frac{3}{2}} K_0  > 0,\\
 & 2 K_0^2 - K_a K_a =0,\\
 &\det(\twomat{K}) =0.
 \end{split}
 \end{equation}
 
 $\twomat{K}=0$ if and only if
 \begin{equation} \label{A.15}
 \begin{split}
 &\trace(\twomat{K}) = \sqrt{\frac{3}{2}} K_0  = 0,\\
 & 2 K_0^2 - K_a K_a =0,\\
 &\det(\twomat{K}) =0.
 \end{split}
 \end{equation}
 \end{theorem}
 
 With this theorem we have expressed the properties of the matrix $\twomat{K}$
 in terms of the expansion coefficients $K_\alpha$, $\alpha=0,\ldots,8$.
 The conditions explicitly written in terms of $K_0$ and $K_a$ in \eqref{A.12}
 to \eqref{A.15} are of the type of {\em light-cone} conditions familiar
 from the two-Higgs-doublet model; see~(36) of~\cite{Maniatis:2006fs}.
 But the determinant condition, trilinear in $K_\alpha$, is specific for the 3HDM.

 To express also $\det(\twomat{K})$ in terms of the expansion coefficients~$K_\alpha$,
 $\alpha=0,\ldots,8$, we proceed as follows (see also~\cite{Ivanov:2010ww}). We introduce,
 along with the matrix~$\twomat{K}$,
 a matrix~$\twomat{M}=(M_{ij})$:
 \begin{equation} \label{A.16}
 M_{ij} = \epsilon_{ikl} \epsilon_{jmn} K_{mk} K_{nl}.
 \end{equation}
 For a hermitian matrix~$\twomat{K}$ also $\twomat{M}$ is hermitian.
 For any $U \in U(3)$ we have the relation
 \begin{equation} \label{A.17}
 \epsilon_{ijk} U_{i i'} U_{j j'} U_{k k'} = \epsilon_{i' j' k'} \det(U).
 \end{equation}
 Using this we find easily that under a transformation~\eqref{3.2a} 
 of~$\twomat{K}$ we get also for $\twomat{M}$
 \begin{equation} \label{A.18}
 \twomat{M}' = U \; \twomat{M} \; U^\dagger.
 \end{equation}
 Furthermore we find
 \begin{equation} \label{A.19}
 \det( \twomat{K} ) = \frac{1}{3!} \trace( \twomat{K} \twomat{M} ).
 \end{equation}
 Consider now a unitary transformation~$U$ which diagonalises $\twomat{K}$; 
 see \eqref{A.1}.
 
 We find then from~\eqref{A.16}
 \begin{equation} \label{A.20}
 U \twomat{M} U^\dagger =
 \begin{pmatrix}
 2\kappa_2 \kappa_3 & 0 & 0 \\
 0 & 2\kappa_1 \kappa_3 & 0 \\
 0 & 0 & 2\kappa_1 \kappa_2
 \end{pmatrix},
 \end{equation}
 and
 \begin{equation} \label{A.21}
 \det( \twomat{K}) = \frac{1}{3!} \trace( \twomat{K} \twomat{M} ) = \kappa_1 \kappa_2 \kappa_3,
 \end{equation}
 \begin{equation} \label{A.22}
 \trace(\twomat{M}) = (\trace(\twomat{K}))^2 - \trace(\twomat{K}^2).
 \end{equation}
 As for $\twomat{K}$ in \eqref{2.5} we can expand $\twomat{M}$ in terms
 of $\lambda_\alpha$,
 \begin{equation} \label{A.23}
 \twomat{M} = \frac{1}{2} M_\alpha \lambda_\alpha,\qquad
 M_\alpha = \trace(\twomat{M} \lambda_\alpha).
 \end{equation}
 Inserting here \eqref{A.16} we get the expression of $M_\alpha$ in terms
 of the $K_\beta$ \eqref{2.6} as follows:
 \begin{equation} \label{A.24}
 M_\alpha = G_{\alpha \beta \gamma} K_\beta K_\gamma
 \end{equation}
 where 
 \begin{multline} \label{A.25}
 G_{\alpha \beta \gamma}  = \frac{1}{4} 
 \bigg\{
 \trace(\lambda_\alpha) \trace(\lambda_\beta) \trace(\lambda_\gamma)
 + \trace(\lambda_\alpha \lambda_\beta \lambda_\gamma + \lambda_\alpha \lambda_\gamma \lambda_\beta)
 - \trace(\lambda_\alpha) \trace(\lambda_\beta \lambda_\gamma)\\
 - \trace(\lambda_\beta) \trace(\lambda_\gamma \lambda_\alpha)
 - \trace(\lambda_\gamma) \trace(\lambda_\alpha \lambda_\beta)
 \bigg\}.
 \end{multline}
 Clearly, $ G_{\alpha \beta \gamma}$ is completely symmetric in
 $\alpha$, $\beta$, $\gamma$.
 Explicitly we get
 \begin{equation} \label{A.26}
 G_{0 \beta \gamma} = \sqrt{\frac{3}{2}} \delta_{\beta 0} \delta_{\gamma 0}
 - \frac{1}{\sqrt{6}} \delta_{\beta \gamma},\qquad
 G_{ a b c} = d_{a b c}
 \end{equation}
 with $d_{a b c}$ the usual symmetric constants of $SU(3)$; see, for instance,
 appendix~C of~\cite{Nachtmann}.
 From \eqref{A.19}, \eqref{A.23}, and \eqref{A.24} we find
 \begin{equation} \label{A.27}
 \det{\twomat{K}} = \frac{1}{12} K_\alpha M_\alpha =
 \frac{1}{12} G_{\alpha \beta \gamma} K_\alpha K_\beta K_\gamma.
 \end{equation}
 This is the desired expression of $\det(\twomat{K})$ in terms of
 the $K_\alpha$.
 
 Finally we discuss the transformation from the basis
 $\alpha=0,1,\ldots,7,8$ to $+,1,\ldots,7,-$.
 This is achieved by the matrix
 \begin{equation} \label{A33}
 S =
 \begin{pmatrix}
 \sqrt{\frac{2}{3}} & 0 & \sqrt{\frac{1}{3}}\\
 0 & \unitmatrix_7 & 0 \\
 \sqrt{\frac{1}{3}} & 0 & -\sqrt{\frac{2}{3}}
 \end{pmatrix}
 =
 \begin{pmatrix}
 S_{+0} & 0 & S_{+8}\\
 0 & \unitmatrix_7 & 0 \\
 S_{-0} & 0 & S_{-8}
 \end{pmatrix} .
\end{equation}
We have then, in accord with \eqref{2.10a},
\begin{equation} \label{A33a}
\begin{split}
&K_+= S_{+0} K_0 + S_{+8} K_8 = \sqrt{\frac{2}{3}} K_0 + \sqrt{\frac{1}{3}} K_8,\\
&K_-= S_{-0} K_0 + S_{-8} K_8 = \sqrt{\frac{1}{3}} K_0 - \sqrt{\frac{2}{3}} K_8,\\
&K_a, \quad a=1,\ldots,7 \quad \text{unchanged.}
\end{split}
\end{equation}
The matrix $S$ satisfies
\begin{equation} \label{A34}
S S^\trans = \unitmatrix_9, \qquad S = S^\trans .
\end{equation}
The basis change for $(\xi_\alpha)$ and $\tilde{E}= (\tilde{E}_{\alpha \beta})$
is then done in an analogous way:
\begin{equation} \label{A35}
\begin{pmatrix} \xi_+ \\ \xi_a \\ \xi_- \end{pmatrix}
=
S
\begin{pmatrix} \xi_0 \\ \xi_a \\ \xi_8 \end{pmatrix}, \quad
\begin{pmatrix}
\tilde{E}_{++} & \tilde{E}_{+b} & \tilde{E}_{+-} \\
\tilde{E}_{a+} & \tilde{E}_{ab} & \tilde{E}_{a-} \\
\tilde{E}_{-+} & \tilde{E}_{-b} & \tilde{E}_{--}
\end{pmatrix}
= S (\tilde{E}_{\alpha \beta} ) S^\trans,
\quad a,b \in \{1,...,7\}, \quad \alpha, \beta \in \{0,\ldots,8\}.
\end{equation}
We have, due to \eqref{A34}, for instance,
\begin{equation} \label{A36}
\xi_\alpha K_\alpha = \xi_+ K_+  + \sum_{a=1}^7 \xi_a K_a + \xi_- K_-.
\end{equation}

  
\section{Example of a 3HDM Higgs potential}
\label{appendixB}

Let us apply the developed formalism to a non-trivial
3HDM potential.
We emphasize that any specific 3HDM can be treated along the following lines.
We will apply the homotopy continuation approach to solve the
systems of polynomial equations allowing us to discuss 
stability and the stationarity points of the model. Of course,
other methods may be applied, like the Groebner-basis approach.
These methods were successfully applied to Higgs potentials
in the past; see for instance~\cite{Maniatis:2006jd,Maniatis:2012ex}.
In these works brief introductions to Groebner-bases and 
homotopy continuation can also be found.

The model we want to study was presented in~\cite{Grimus:2008dr} 
and is based on a $O(2) \times \mathbbm{Z}_2$ symmetry  
involving three Higgs-boson doublets. 
All the elementary particles and in particular the three
Higgs-boson doublets are assigned to
irreducible representations of the $O(2) \times \mathbbm{Z}_2$
symmetry. For the three Higgs-boson doublets
the assignments were chosen as given in Table~\ref{assign}.

\begin{center}
\begin{table}[ht]
\begin{tabular}{c|ccc}
 & $s$ & $U(1)$ & $\mathbbm{Z}_2$ \\
\hline
$\begin{pmatrix} \varphi_1 \\ \varphi_2 \end{pmatrix}$ &
$\begin{pmatrix} 0 & 1 \\ 1 & 0 \end{pmatrix}$ &
$\begin{pmatrix} e^{2i\theta} & 0 \\ 0 & e^{-2i\theta} \end{pmatrix}$ &
$\unitmatrix_2$ \\
$\varphi_3$ & 1 & 1 & -1
\end{tabular}
\caption{\label{assign}Assignments of the transformation behaviour of the Higgs-boson doublets
under the symmetries $s$, $U(1)$, $\mathbbm{Z}_2$.}
\end{table}
\end{center}
Here, the group~$O(2)$ is decomposed into
unitary rotations $U(1)$ and reflections $s$.

The general 3HDM Higgs potential, symmetric under $O(2)\times \mathbbm{Z}_2$
except for the term proportional to $\mu_m$
reads
\begin{multline} \label{V02}
V_{O(2)\times \mathbbm{Z}_2} = \mu_0 \varphi_3^\dagger \varphi_3 
+\mu_{12} \left( \varphi_1^\dagger \varphi_1 + \varphi_2^\dagger \varphi_2 \right)
+\mu_m \left( \varphi_1^\dagger \varphi_2 + \varphi_2^\dagger \varphi_1 \right)\\
+a_1 (\varphi_3^\dagger \varphi_3)^2
+a_2 \varphi_3^\dagger \varphi_3
 \left( \varphi_1^\dagger \varphi_1 + \varphi_2^\dagger \varphi_2 \right) 
+a_3 \left( \varphi_3^\dagger \varphi_1 \cdot \varphi_1^\dagger \varphi_3 +
	\varphi_3^\dagger \varphi_2 \cdot \varphi_2^\dagger \varphi_3  \right) 
+a_4 \varphi_3^\dagger \varphi_1 \cdot \varphi_3^\dagger \varphi_2
+a_4^* \varphi_1^\dagger \varphi_3 \cdot \varphi_2^\dagger \varphi_3 \\
+a_5 \left( (\varphi_1^\dagger \varphi_1)^2 + (\varphi_2^\dagger \varphi_2)^2 \right)
+a_6 \varphi_1^\dagger \varphi_1 \cdot \varphi_2^\dagger \varphi_2
+a_7 \varphi_1^\dagger \varphi_2 \cdot \varphi_2^\dagger \varphi_1 .\hspace{5cm} 
\end{multline}
The term 
$\mu_m \left( \varphi_1^\dagger \varphi_2 + \varphi_2^\dagger \varphi_1 \right)$
breaks the $U(1)$ symmetry softly (for details we refer to~\cite{Grimus:2008dr}).
This model has nine real parameters and one complex parameter~$a_4$, corresponding
to eleven real parameters in total.

With the help of~\eqref{phiK} we write the potential in terms of bilinears. 
We identify the parameters of the potential~\eqref{V02}, but written in the form~\eqref{eq-vdef},
as
\begin{multline} \label{paraK}
\xi_0 = \frac{1}{\sqrt{6}} (\mu_0 + 2 \mu_{12}), \quad
\tvec{\xi} = \left(\mu_m, 0, 0, 0, 0, 0, 0, \frac{1}{\sqrt{3}}(\mu_{12}-\mu_0)\right)^\trans, \quad
\eta_{00} = \frac{1}{6} (a_1 + 2 a_2 + 2 a_5 + a_6),\\
\tvec{\eta} = \left( 0, 0, 0, 0, 0, 0 ,0, \frac{\sqrt{2}}{6} (-a_1-a_2/2+a_5+a_6/2) \right)^\trans, \\
E = \frac{1}{4}
\begin{pmatrix}
a_7 & 0 & 0 & 0 & 0 & 0 & 0 & 0\\
0 & a_7 & 0 & 0 & 0 & 0 & 0 & 0\\
0 & 0 & 2 a_5 - a_6 & 0 & 0 & 0 & 0 & 0\\
0 & 0 & 0 & a_3 & 0 & \re (a_4) & \im (a_4) & 0\\
0 & 0 & 0 & 0 & a_3 & \im (a_4) & -\re (a_4) & 0\\
0 & 0 & 0 & \re (a_4) & \im (a_4) & a_3 & 0 & 0\\
0 & 0 & 0 & \im (a_4) & -\re (a_4) & 0 & a_3 & 0\\
0 & 0 & 0 & 0 & 0 & 0 & 0 & 4/3 a_1 - 4/3 a_2+ 2/3 a_5+ 1/3 a_6
\end{pmatrix}.
\end{multline}
Obviously, all parameters are real in terms of bilinears.

We choose as an explicit numerical 
example the following values for the parameters, where we take
only the quartic couplings from the 
reference point in~\cite{Grimus:2008dr}:
\begin{equation} \label{para}
\begin{split}
&a_1=2.5, \quad
a_2=3, \quad
a_3=-5, \quad
a_4=-0.0474041, \quad
a_5=1.5, \quad
a_6=2, \quad
a_7=3, \\
&\mu_0 = - 90,774~\text{GeV}^2, \quad
\mu_{12} = - 75,645~\text{GeV}^2, \quad
\mu_m = - 45,387~\text{GeV}^2. \\
\end{split}
\end{equation}
Actually, we start with all quartic parameters as given 
in \eqref{para} and then fix the quadratic parameters 
by the condition of a vanishing gradient of the potential,
employing $K_{v 0} = v_0^2/\sqrt{6}$ (see \eqref{5.6} with
$v_0$ as given in~\eqref{5.5}).
In this way we ensure that there is at least one stationary solution
which corresponds to the correct vacuum expectation value. 
Let us note that this procedure by no means guarantees that
the corresponding potential is stable and has 
a global minimum with the correct partially broken electroweak
symmetry -- as we will see below.

The stability and stationarity equations are polynomial systems of equations
as given in~\eqref{4.7}, \eqref{4.10} and \eqref{5.12}, \eqref{6.6}, respectively.
In this example we apply for all the polynomial systems of equations 
the homotopy continuation approach as implemented in the 
PHCpack package~\cite{phcpack}. We first look for solutions disregarding the
inequalities and then select by hand all solutions which fulfill them.
Technically, for the real indeterminats we take into account solutions with an imaginary 
part smaller than 0.001. With respect to 
our computations we remark that for the most
involved cases of systems of equations we encounter 
about a minute of time consumption on an ordinary PC.

We start with studying stability of the potential; see section~\ref{stability}.
To this end we separate the quadratic and the quartic terms of the
potential. Inserting the parameters~\eqref{paraK} 
into \eqref{eq-vk}, \eqref{eq-vk4}, yields
\begin{equation} \label{J24}
\begin{split}
J_2(\tvec{k}) = &\frac{\mu_0 + 2 \mu_{12}}{\sqrt{6}} 
+ \left( \frac{\mu_{12} - \mu_0}{\sqrt{3}} \right) k_8
+ \mu_m k_1, \\
J_4(\tvec{k}) = & \frac{1}{6}(a_1 + 2 a_2 + 2 a_5 + a_6)
+ \frac{1}{3 \sqrt{2}} (-2 a_1 - a_2 + 2 a_5 + a_6) k_8
+ \frac{a_7}{4} (k_1^2+k_2^2)
+ \frac{1}{4} (2 a_5 - a_6 ) k_3^2 \\
&
+ \frac{a_3}{4} (k_4^2+k_5^2+k_6^2+k_7^2)
+ \frac{\re (a_4)}{2} (k_4 k_6 - k_5 k_7)
+ \frac{\im (a_4)}{2} (k_4 k_7 + k_5 k_6)
+ \frac{1}{12}(4 a_1 - 4 a_2 + 2 a_5 + a_6) k_8^2
\end{split}
\end{equation}
with the parameters given in~\eqref{para}.
Now we have to find the stationary points of $J_4(\tvec{k})$,
that is, we have to solve the systems of equations~\eqref{4.7} and
\eqref{4.10}, respectively. 

In case of~\eqref{4.7}
the invariants are the eight components of the vector~$\tvec{k}$ and one
Lagrange multiplier~$u$. 
in case of~\eqref{4.10} the invariants are the vector components of
$\tvec{w}$, $\tvec{w}^\dagger$, 
(the eigenvector of the matrix $\twomat{K}$, see \eqref{A10c}),
as well as one Lagrange multiplier $u$.
We decompose the vector components of $\tvec{w}$, $\tvec{w}^\dagger$ into
real and imaginary parts such that in this form all
indeterminants, that is, $\re (w_1)$, $\im(w_1)$,  $\re (w_2)$, $\im (w_2)$, $\re (w_3)$, $\im (w_3)$, $u$,
are real. Then we split the system of equations~\eqref{4.10} into its real
and imaginary parts. Note that the last equation $\tvec{w}^\dagger \tvec{w}-1=0$
has only a trivial imaginary part. Eventually, we end up with seven equations in
seven indeterminants in this case. 

With respect to the system~\eqref{4.7} we detect four solutions $\tvec{k}$ and~$u$
fulfilling the inequality $2-\tvec{k}^2>0$.
Plugging these solutions into $J_4(\tvec{k})$ in~\eqref{J24} we find solely positive values.
With respect to the systems~\eqref{4.10} we find solely solutions
which give, with respect to $J_4(\tvec{k})$, also positive values.
Stability in general requires that there is 
no stationary direction~$\tvec{k}$ with $J_4(\tvec{k})<0$
or $J_4(\tvec{k})=0$ but $J_2(\tvec{k})<0$. In our example
$J_4(\tvec{k})$ is positive for all stationary directions $\tvec{k}$, therefore,
the potential is stable in the strong sense for the chosen parameters.

Since the potential with parameters~\eqref{para} is stable we proceed by
studying the stationary points; see section~\ref{stationarity}. 
To this end we plug the parameters \eqref{paraK} into
the potential~\eqref{eq-vdef}. This gives
\begin{equation}
\begin{split}
V =& 
\frac{\mu_0 + 2 \mu_{12}}{\sqrt{6}} K_0
+ \left( \frac{\mu_{12} - \mu_0}{\sqrt{3}} \right) K_8
+ \mu_m K_1\\
& + 
\frac{1}{6}(a_1 + 2 a_2 + 2 a_5 + a_6) K_0^2
+ \frac{a_7}{4} K_1^2
+ \frac{a_7}{4} K_2^2
+ \frac{1}{4} (2 a_5 - a_6 ) K_3^2
+ \frac{a_3}{4} (K_4^2+K_5^2+K_6^2+K_7^2) \\
&
+ \frac{\re (a_4)}{2} (K_4 K_6 - K_5 K_7)
+ \frac{\im (a_4)}{2} (K_4 K_7 + K_5 K_6)
+ \frac{1}{3 \sqrt{2}} (-2 a_1 - a_2 + 2 a_5 + a_6) K_0 K_8\\
&
+ \frac{1}{12}(4 a_1 - 4 a_2 + 2 a_5 + a_6) K_8^2 .
\end{split}
\end{equation}
Now we have to solve the
systems of polynomial equations~\eqref{5.12} (corresponding to
solutions which break electroweak symmetry fully) and~\eqref{6.6}
(corresponding to solutions with break electroweak symmetry partially leaving
 the electromagnetic $U(1)$ symmetry intact).

For the set of equality equations~\eqref{5.12} we
find one real solution 
fulfilling the inequalities $2K_0^2 - K_a K_a>0$ and $K_0>0$,
\begin{equation} \label{solFB}
K_0 = 24,705.6~\text{GeV}^2 ,\;
K_1 =  30,258~\text{GeV}^2 ,\;
K_8 = 17,469.5~\text{GeV}^2 ,\;
K_{2/3/4/5/6/7} = 0.
\end{equation}
This solution corresponds to a potential value
of $V(K) = - 1.83109\cdot 10^{9} \text{ GeV}^4$.
Since this solution originates from the set~\eqref{5.12} 
it corresponds to a stationary point with fully broken
electroweak~symmetry.

Eventually, we write the set~\eqref{6.6}
again with explicit real and imaginary parts of the vectors 
$\tvec{w}$, $\tvec{w}^\dagger$ and subsequently separate
real and imaginary parts of the equations. We have
here eight equations in eight indeterminants,
$K_0$, $\re (w_1)$, $\im(w_1)$,  $\re (w_2)$, $\im (w_2)$, $\re (w_3)$, $\im (w_3)$, $u$.
In terms of bilinears, employing \eqref{A10c} we encounter
three real solutions
fulfilling the inequality $K_0>0$:
\begin{small}
\begin{equation} \label{solPB}
\begin{split}
&K_0 = 14,823.3~\text{GeV}^2 ,\;
K_8 = -20,963.4~\text{GeV}^2 ,\;
K_{1/\ldots/7} = 0, \\
&K_0 = 39,367.3~\text{GeV}^2 ,\;
K_1 = -21,327.1~\text{GeV}^2, \;
K_5 = \pm 33,865.6~\text{GeV}^2, \;
K_7 = - K_5, \;
K_8 = -18,734.2~\text{GeV}^2, \;
K_{2/3/4/6} = 0.
\end{split}
\end{equation}
\end{small}
The first solution in \eqref{solPB} corresponds
to a potential value of $V(K) = -8.24\cdot 10^{8} \text{ GeV}^4$
and the other two solutions to
$V(K) = -1.54\cdot 10^{9} \text{ GeV}^4$.
These solutions correspond to the correct electroweak symmetry breaking.

In addition, we always have the trivial solution with vanishing
bilinears, corresponding to a vanishing potential. This solution
corresponds to an unbroken electroweak symmetry.

The global minimum is given by the stationary point corresponding
to the deepest potential value. In this example 
the deepest stationary point is given by~\eqref{solFB} and corresponds to 
a fully broken electroweak symmetry which is physically
not acceptable.

Our analysis clearly shows that requiring the potential to have a stationary
point giving the desired electroweak 
symmetry breaking and vacuum expectation value does not 
guarantee that one has a viable model.
In contrast, the study of stability and {\em all} stationary points reveals where the 
true global minimum of the potential is. Then one has to check if,
at this global minimum, one has the desired partial electroweak
symmetry breaking. As we have seen, our methods employing
bilinears allow to perform these investigations in an efficient way.


\end{document}